\documentclass[twocolumn,tighten,times]{aastex63}
\usepackage{lineno}
\pdfoutput=1
\usepackage[utf8]{inputenc} 
\usepackage{amsmath} 

\usepackage{graphicx}
\usepackage{xspace}
\usepackage{multirow}
\usepackage{ulem}
\usepackage{xfrac}

\usepackage{enumitem}
\setlist{leftmargin=*}
\usepackage[figure, figure*]{hypcap}

\newcommand\planck{{\sl Planck}\xspace}
\newcommand\herschel{{\sl Herschel}\xspace}

\newcommand\msun{\ensuremath{{\rm M}_\sun}\xspace}

\newcommand\dd{\mathrm{d}}

\newcommand{\pc}{\ensuremath{\mathrm{pc}}}

\newcommand{\kms}{\ensuremath{\mathrm{km\,s^{-1}}}}
\newcommand{\Kkms}{\ensuremath{\mathrm{K\ \kms}}}
\newcommand{\magn}{\ensuremath{\mathrm{mag}}}
\let\kkms\Kkms

\newcommand{\A}[1]{\ensuremath{A_{\mathrm{#1}}}\xspace}
\newcommand{\ak}{\ensuremath{A_{\rm K}}\xspace}
\newcommand{\av}{\ensuremath{A_{\rm V}}\xspace}

\newcommand{\W}[1]{\ensuremath{{W_{\rm #1}}}\xspace}
\newcommand{\X}[1]{\ensuremath{{X({\rm #1})}}\xspace}
\newcommand{\N}[1]{\ensuremath{{\rm {N}({#1})}}\xspace}
\newcommand{\E}[1]{\ensuremath{\times10^{#1}}\xspace}
\let\e\E

\newcommand{\HI}{\ensuremath{{\rm H\:\!\text{\small I}}}\xspace}
\newcommand{\htwo}{\ensuremath{{\rm H}_2}\xspace}
\newcommand{\nhtotal}{\ensuremath{\mathrm{\left[\N{\HI} + 2 \N{\htwo}\right]}}\xspace}

\newcommand{\taudust}{\ensuremath{\tau_{353}}\xspace}

\newcommand{\xco}{\ensuremath{X_{\rm CO}}\xspace}
\newcommand{\wco}{\ensuremath{W_{\rm CO}}\xspace}
\newcommand{\tdust}{\ensuremath{T_{\rm dust}}\xspace}
\newcommand{\aco}{\ensuremath{\alpha_{\rm CO}}\xspace}

\newcommand{\xfunit}{\ensuremath{{\rm cm}^{-2}\ {(\Kkms)}^{-1}}\xspace}
\newcommand{\msunpc}{\ensuremath{\ \msun\, \pc^{-2}}\xspace}
\newcommand{\dgunit}[1][]{\ensuremath{\mathrm{\ cm^{-2}\, \magn^{-1}}}\xspace}
\newcommand{\alphaunit}{\ensuremath{\msun/(\Kkms\ \pc^2)}\xspace}
\newcommand{\mdgunit}[1]{\ensuremath{\mathrm{\ g\ cm^{-2}\ \magn^{-1}}}\xspace}
\newcommand{\twelveco}{\ensuremath{{}^{12}\text{CO}}\xspace}

\newcounter{paranum}

\newcommand{\akmax}{\ensuremath{A_{\rm K,max}}\xspace}
\newcommand{\akmin}{\ensuremath{A_{\rm K,min}}\xspace}

\newcommand{\avgsig}{\ensuremath{\langle\Sigma\rangle}}
\newcommand{\sigcon}{\ensuremath{R_{\rm dg}^\prime\xspace}}

\newcommand{\M}[1]{\ensuremath{{ {M}_{\rm #1}}}\xspace}
\newcommand{\R}[1]{\ensuremath{{ {R}_{\rm #1}}}\xspace}

\newcommand{\avg}[1]{\ensuremath{\langle#1\rangle}\xspace}

\newcommand{\glon}{\ensuremath{l}\xspace}
\newcommand{\glat}{\ensuremath{b}\xspace}

\defcitealias{2001ApJ...547..792D}{DHT}
\defcitealias{2020ApJ...898....3L}{LD20}
\defcitealias{2020MNRAS.493..351C}{C+20}
\defcitealias{2010A&A...519L...7L}{LAL10}

\submitjournal{ApJ}

\begin{document}

\shorttitle{Dust and CO in Nearby GMCs}
\shortauthors{Lewis et al.}


\title{Systematic Investigation of Dust and Gaseous CO in 12 Nearby Molecular Clouds}

\correspondingauthor{Charles J. Lada}
\email{clada@cfa.harvard.edu}

\author[0000-0001-5199-3522]{John Arban Lewis}
\affiliation{Center for Astrophysics | Harvard \& Smithsonian \\
60 Garden St \\
Cambridge, MA 02138}

\author[0000-0002-4658-7017]{Charles J. Lada}
\affiliation{Center for Astrophysics | Harvard \& Smithsonian \\
60 Garden St \\
Cambridge, MA 02138}

\author[0000-0003-0109-2392]{T. M. Dame}
\affiliation{Center for Astrophysics | Harvard \& Smithsonian \\
60 Garden St \\
Cambridge, MA 02138}

\begin{abstract}

We report on the first uniform and systematic study of dust and molecular gas in nearby molecular clouds. We use surveys of dust extinction and emission to determine the opacity and map the distribution of the dust within a dozen local clouds in order to derive a uniform set of basic cloud properties.  We find: (1) the average dust opacity $\langle\kappa_{d,353}\rangle  = 0.8\ {\rm cm^{2}\, g^{-1}}$ with variations of a factor of $\sim$ 2 between clouds, (2) cloud probability density functions are exquisitely described by steeply falling power-laws with a narrow range of slope, and (3)  a tight $M_{\rm GMC} \sim R_{\rm GMC}^2$  scaling relation for the cloud sample, indicative of a cloud population with an exactingly constant average surface density above a common fixed boundary. We compare these results to uniformly analyzed CO surveys. We measure the CO mass conversion factors and assess the efficacy of CO for tracing the physical properties of molecular clouds. We find $\langle \alpha_{\rm CO}\rangle = 4.31 \pm 0.67$ $M_\odot$ (K km s$^{-1}$ pc$^2$)$^{-1}$ (corresponding to \xco = 1.97 $\times$ 10$^{20}$ cm$^{-2}$(K km s$^{-1}$)$^{-1}$). We demonstrate that CO observations are a poor tracer of column density and structure on sub-cloud spatial scales. On cloud scales,  CO observations can provide measurements consistent with those of the dust, provided data are analyzed in a similar, self-consistent fashion. Measurements of average giant molecular cloud surface density are sensitive to choice of cloud boundary.  Care must be exercised to adopt common fixed boundaries when comparing surface densities for cloud populations within and between galaxies.

\end{abstract}


\section{Introduction}\label{sec:intro}
The discovery of CO in the interstellar medium (ISM) a little more than half a century ago ushered in a compelling exploration of the cold ($T_K \approx 10$~K) cosmos. 
It is in the cold component of the ISM that nearly all star formation has occurred since the earliest epochs of cosmic history. Giant molecular clouds (GMCs) are the primary component of the cold ISM and the principal sites of star formation in galaxies. As such, they play a key role in galaxy evolution. In the Milky Way, GMCs have been extensively studied using both observations of CO \citep[e.g.,][]{2015ARA&A..53..583H} and dust \citep[e.g.,][]{2007prpl.conf....3L}. 
Analysis of both global CO surveys \citep[e.g.,][]{1987ApJ...319..730S,2001ApJ...547..792D,2010ApJ...723..492R,2013ApJS..209....8D,2016ApJ...822...52R,2017ApJ...834...57M}
and individual case studies \citep[e.g.,][]{1976ApJS...32..603L,  1977ApJ...215..521K,1978ApJ...226L..39L,1980ApJ...241..676B,1994ApJ...425..641S,2008ApJS..177..341N,2008ApJ...679..481P,2010ApJS..191..232B,2011ApJS..196...18B,2016ApJS..226...13B}
have provided important information on the basic physical properties, including the sizes, masses, temperatures, and dynamics of GMCs throughout the Milky Way. Molecular clouds are primarily composed of molecular hydrogen, which is better traced by dust than by CO \citep{2009ApJ...692...91G}, making observations of the dust especially powerful probes of GMCs. This is because dust is not only well mixed with molecular hydrogen but also is more abundant than CO; and unlike CO, its emission is optically thin. However, observations of dust provide no information on the kinematics or dynamics of the clouds. Moreover, unlike CO measurements, observations of dust in GMCs are mostly limited to local regions of the Milky Way. Indeed, observations of dust from individual GMCs are extremely difficult, if not impossible, to obtain outside the Milky Way in other than the nearest external galaxies, e.g., LMC \citep{2010MNRAS.406.2065H} and Andromeda/M31 \citep{2020ApJ...890...42F,2021ApJ...912...68V}.  

Nonetheless, detailed studies of local GMCs have been conducted using both dust extinction and emission data. These investigations  have provided more detailed and robust measurements of GMC sizes, masses, and internal structures than CO observations of the same regions 
\citep[e.g.,][]{1994ApJ...429..694L,2001Natur.409..159A,2007A&A...462L..17A,2014A&A...566A..45L,2018A&A...620A..24H,2021ApJ...919...35Z}. One of the most systematic surveys for dust extinction and emission in local GMCs has been presented in a series of papers by Lombardi and collaborators \citep[e.g.,][]{2006A&A...454..781L,2008A&A...489..143L,2010A&A...512A..67L,2011A&A...535A..16L,2014A&A...566A..45L,2014A&A...565A..18A,2009ApJ...703...52L,2016A&A...587A.106Z,2017A&A...606A.100L}. 
In particular, this series of papers provided a systematic study of a sample of 11 local GMCs and produced a uniform set of basic cloud properties as well as uniform measurements of the internal structures of the clouds. Analysis of this local GMC sample established that GMCs have a stratified internal structure and that their probability distribution functions (PDFs) were best described by simple power-law functions across essentially the entire span of cloud extinction above the completion limits of the observations. It was also found that these GMCs display a very tight power-law scaling between their sizes and masses with an index (2) indicating that the clouds are characterized by the same average surface density to a very high degree of precision ($\sim$ 10\%). This latter finding provided secure confirmation of an important scaling law originally identified by \citet{1981MNRAS.194..809L} from early CO observations and has broad implications. For example, these  results indicate that GMCs cannot obey a Kennicutt--Schmidt scaling law \citep{2013ApJ...778..133L}.  Comparison of this survey data with catalogs of young stellar objects in the same clouds, resulted in the construction of star formation laws for the GMC sample that indicated the importance of dense gas for setting local star formation rates \citep{2010ApJ...724..687L} and that could be continuously extrapolated to match the star formation law characterizing external galaxies \citep{2012ApJ...745..190L}.

Because CO is still the primary tracer of molecular gas in the universe and is likely to remain so for some time to come, understanding the relation between gas and dust in GMCs is critical for evaluating the efficacy of CO for measuring basic GMC properties in the more distant regions of the Milky Way and especially in external galaxies. Attaining such an understanding is also important for ultimately constructing a more detailed picture of the internal process of star formation, itself. Surprisingly, there have been only a few direct comparisons between detailed dust and CO mapping surveys in the local GMC sample \citep[e.g.,][]{2006A&A...454..781L,2008ApJ...679..481P,2013MNRAS.431.1296R,2010ApJ...721..686P,2015ApJ...805...58K,2021ApJ...908...76L}. The purpose of this paper is to improve this situation. Here, we identify and systematically investigate in some detail the relationship between dust and CO emission in a new sample  of local GMCs. This new sample consists of 12 GMCs and though it has significant overlap with the \citet{2010A&A...519L...7L} sample, it  also includes clouds from a wider volume of the local Milky Way. To produce a meaningful comparison of the gas and dust measurements, we use the CO survey of \citet*{2001ApJ...547..792D} to provide uniform measurements of CO emission in the clouds and extinction calibrated \planck data to provide a uniform set of dust maps of the sample clouds.

We employ the dust observations to produce relatively robust determinations of the basic physical properties of the clouds, including dust properties and cloud masses, sizes and internal structures. We perform a similar study of the CO data and make a detailed comparison of the results with those from the dust analysis in an attempt to evaluate the efficacy of CO observations for determining the masses, sizes, surface densities, and internal structures of molecular clouds. In Section \ref{sec:data}, we describe our cloud sample and the CO and dust observations we use in our analysis. In Section \ref{sec:gamma}, we compare the dust extinction and emission observations to both calibrate the \planck optical depths and measure the dust opacity and its cloud-to-cloud variation.  In Section \ref{sec:cointro}, we will derive the global CO conversion factors (\aco and \xco) for the CO emitting gas in each cloud.  In Section \ref{sec:mass_size0}, we investigate the internal structure of the clouds as determined by the dust and CO observations. In Section \ref{sec:mass_size1}, we investigate the mass--size relation and explore its link to the internal structures of the clouds. Section \ref{sec:summary} summarizes our results.

\section{Data}\label{sec:data}
\subsection{CO Observations:}\label{sec:mini}
The \citet*{2001ApJ...547..792D} (hereafter, DHT) survey of \twelveco $J$=$1\rightarrow0$ is composed of the data from 37 separate surveys covering the majority of the Milky Way galaxy. The average beamwidth (FHWM) for the surveys is $\Theta_{\rm beam} = 8\farcm6$. Each \citetalias{2001ApJ...547..792D} survey has a uniform RMS noise, and achieved flat spectral baselines using frequent switching in either position or frequency. All surveys included were sampled on slightly better than every beamwidth spacing except for Ophiuchus and RCrA which were sampled every other beamwidth. Channels without emission are masked using the {\it moment masking} method described in \citet{2011arXiv1101.1499D} which captures even weak emission at the edges of clouds. In practice, we use the moment-masked cubes provide by the online CO survey archive\footnote{Data are available on the online archive at: \url{lweb.cfa.harvard.edu/rtdc/CO/}}.  The average noise in the velocity integrated CO intensity is $\sim0.3\ \kkms$ over a typical integration range of 10 channels.

\begin{figure*}[t]
    \centering
    \includegraphics[width=\textwidth]{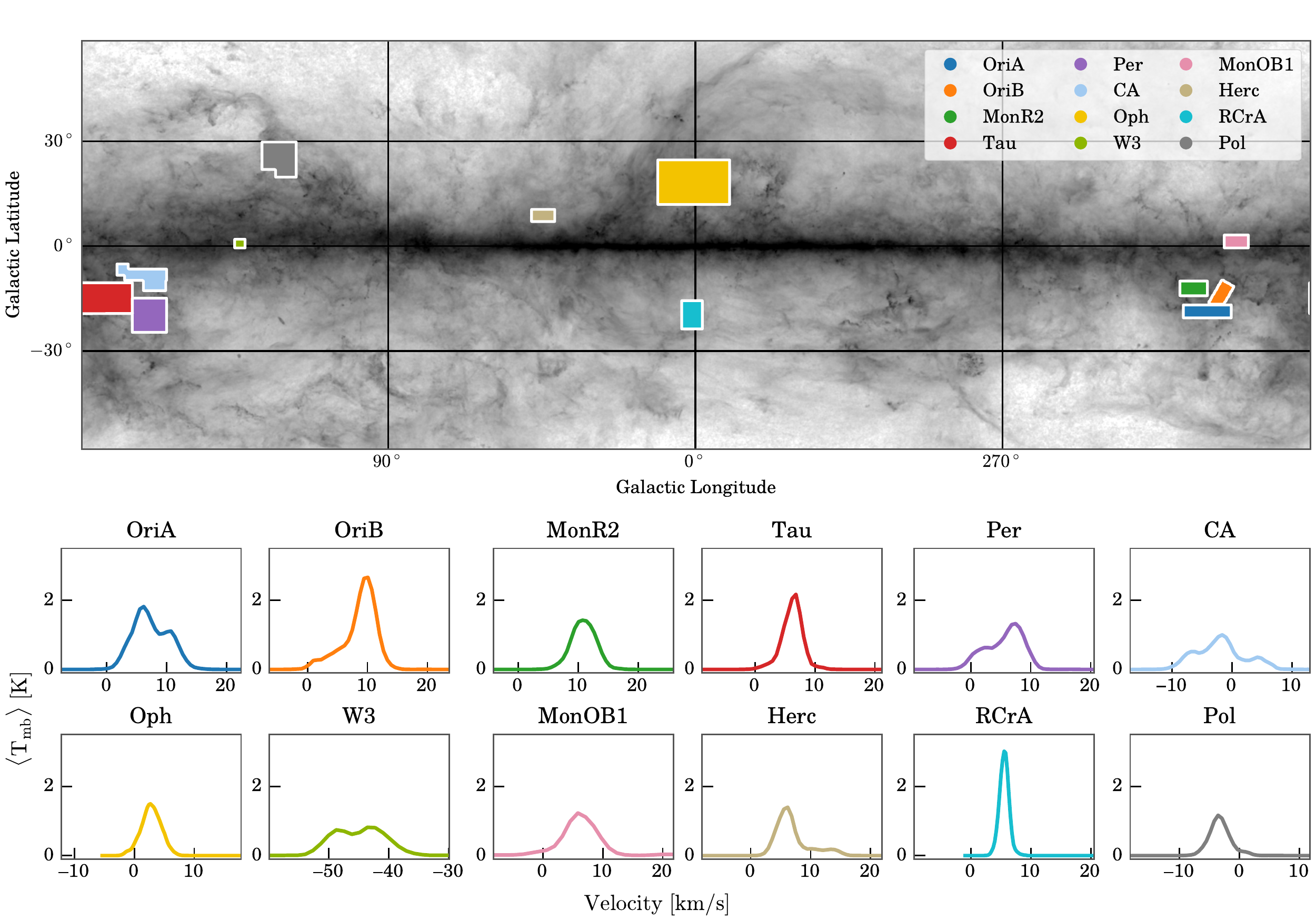}
    \caption{\label{fig:survey} Locations of surveyed clouds and their corresponding average spectra for pixels with $\wco>1.5\ \Kkms$. The colors of the spectra match the colors of the corresponding mask boundaries that define the surveyed regions on the map. The grayscale is the \planck optical depth map. Zoomed in maps of these regions are available in Appendix \ref{app:maps}.}
\end{figure*}

Figure \ref{fig:survey} and Table \ref{tab:regions} show the clouds that were selected for this study and the \citetalias{2001ApJ...547..792D} survey in which the data is found. The spatial resolution at the distance to the clouds spans 0.3-5 pc (average 1.3 pc). { In Table \ref{tab:regions}, t}he galactic longitude and latitude ranges are the full $l$ and $b$ range spanned by the region masks seen in Figure \ref{fig:survey}. { The masks used to define each cloud region can be seen in detail in Appendix \ref{app:maps}.} We chose the regions to be well separated in velocity space, in order to limit confusion and enable comparison with \cite*{2010A&A...519L...7L} (hereafter, LAL10).

W3 has a significant amount of unassociated CO in a foreground velocity component, so we restrict ourselves to a spectral window spanning $-54$ to $-32$ \kms. The foreground component is well separated in velocity and is not included in our analysis; however, there is dust associated with that CO that will contaminate our measurements. We define a CO foreground fraction, $\sfrac{{\rm W(CO)}_{\rm fore}}{{\rm W(CO)}_{\rm total}}$, as a function of $l$ and $b$, and remove the same fraction from the dust extinction and propagate the errors. This was only necessary for W3. For all other sources, \wco was derived by integrating the entire velocity range in the moment-masked cubes from the \citetalias{2001ApJ...547..792D} survey data cubes.

\begin{deluxetable*}{lcccCR}
\tablecaption{Regions}\label{tab:regions}
\tablehead{\colhead{Name} & \colhead{Survey} & \colhead{Distance} & \colhead{$\sigma$\tablenotemark{\footnotesize a}} & \colhead{$l^\circ$} & \colhead{$b^\circ$} \\ 
  &  & \colhead{(pc)} & \colhead{(K)} & \colhead{($^\circ$)} & \colhead{($^\circ$)}}
\startdata
Orion A    &  DHT 27  &   432\tablenotemark{\footnotesize z}  &   0.26 &    217.1  < l < 202.9      &     -21.1  < b < -16.9  \\
Orion B    &  DHT 27  &   423\tablenotemark{\footnotesize z}  &   0.26 &   209.8  < l < 202.2      &     -17.1  < b < -9.6   \\
Mon R2   &  DHT 27  &   778\tablenotemark{\footnotesize z}  &   0.26 &   218.1  < l < 209.9      &     -14.6  < b < -9.9   \\
  Taurus    &  DHT 21  &   153\tablenotemark{\footnotesize l}  &   0.25 &   180.1  < l < 164.9      &     -19.8  < b < -10.4  \\
  Perseus    &  DHT 21  &   240\tablenotemark{\footnotesize l}  &   0.25 &   165.1  < l < 154.9      &     -25.1  < b < -14.9  \\
  California     &  DHT 21  &   450\tablenotemark{\footnotesize l}  &   0.25 &   169.6  < l < 154.9      &     -13.1  < b < -4.9   \\
  Ophiuchus    &  DHT 37  &   144\tablenotemark{\footnotesize z}  &   0.31 &    11.1  < l < 349.9      &     +11.9  < b < +25.1  \\
  W3     &  DHT 17  &  2000\tablenotemark{\footnotesize r}  &   0.31 &   135.1  < l < 131.9      &      -0.6  < b < +2.1   \\
Mon OB1   &  DHT 26  &   745\tablenotemark{\footnotesize z}  &   0.24 &   205.1  < l < 197.9      &      -0.6  < b < +3.3   \\
 Hercules    &  DHT 09  &   227\tablenotemark{\footnotesize z}  &   0.18 &    48.1  < l < 41.2       &      +6.9  < b < +10.8  \\
 R Corona A    &  DHT 03  &   130\tablenotemark{\footnotesize a}  &   0.30 &     4.1  < l < 357.9      &     -24.1  < b < -15.9  \\
  Polaris    &  DHT 16  &   352\tablenotemark{\footnotesize z}  &   0.13 &   127.0  < l < 117.0      &     +19.9  < b < +34.1  \\
\enddata

\tablerefs{Distance references: ({\it z}) \citet{2019ApJ...879..125Z}, ({\it l}) Lombardi et al. (2010)b, ({\it a}) Alves et al. (2014), ({\it r}) \citet{2008hsf2.book.....R}}
\tablenotetext{a}{$\sigma$ is the RMS noise/channel in Kelvin from Table 1 in \citet{2001ApJ...547..792D}.}
\tablecomments{The Galactic latitude and longitude ranges are the maximum extent in ($l$,$b$) of the boundary defined for the clouds. The boundaries can be found in Appendix~\ref{app:maps}. }
\end{deluxetable*}
\nocite{2010A&A...512A..67L,2014A&A...565A..18A}

\subsection{Dust Emission Optical Depth from \planck}\label{sec:planck}
\cite{2014A&A...571A..11P} produced an all-sky, 5\arcmin\ resolution model of the thermal dust emission using \planck 353, 545, and 857 GHz and IRAS 100\micron\ maps. Thermal dust was modeled as a modified blackbody, 
\begin{linenomath}
$$ I_\nu = \tau_{\nu_0} B_\nu(\tdust) \left(\frac{\nu}{\nu_0}\right)^\beta ,$$
\end{linenomath}
where $\nu_0 = 353$\ GHz, $\tau_{\nu_0}$ is the dust optical depth at $\nu_0$, and $\beta$ is the dust opacity law index. This modified blackbody was fit to the observed spectral energy distribution (SED) in a two-step process to reduce the effects of cosmic infrared background anisotropies and the well-known $\beta - \tdust$ degeneracy that has been discussed in the literature \citep[e.g.][]{2009ApJ...696.2234S}. The SED is fit first at 30\arcmin\ resolution and is fit again at the full resolution using the $\beta$ found from the 30\arcmin\ data. The fit returns the dust optical depth at 353 GHz ($\tau_{353}$), the dust $\beta$ (at 30\arcmin\ resolution), and dust temperature (\tdust).  

The \planck maps are provided in {\it healpix} format\footnote{The data are available here: \url{https://irsa.ipac.caltech.edu/data/Planck/release_1/all-sky-maps/index.html}}. For comparison with CO survey, the \textit{healpix} maps are smoothed to an 8\farcm6 FWHM beam, and reprojected to match the CO data using \texttt{mProject} from the {\it montage} package \citep{2010arXiv1005.4454J,2017PASP..129e8006B}.

\subsection{Dust extinction from 2MASS and NICEST}\label{sec:nicest}
We generate maps of {\it K}-band dust extinction (\ak) using the NICEST method \citep{2009A&A...493..735L}. Estimates of extinction using near-infrared color excess methods \citep{1994ApJ...429..694L,2001A&A...377.1023L} rely on measuring the difference between a star's observed and intrinsic colors. This difference is related to the line-of-sight extinction. Intrinsic colors are estimated from a nearby control field of unreddened stars. Extinctions to individual stars are measured and then smoothed with a weighted average to derive pixel-based extinction. 
The fact that stars are not distributed with uniform density introduces a bias during the smoothing process. Physically, unresolved high extinction substructure reduces the number of background stars relative to what would be expected at the average measured extinction. The NICEST method improves upon this by modifying the measurement weights using the {\it K}-band luminosity function of control field stars to account for the lack of stars in the science field, increasing the extinction estimate.

Extinction maps can be readily created from Two Micron All Sky Survey (2MASS) {\it J}, {\it H}, and {\it K} colors using the online iNICEST tool\footnote{iNICEST: \url{http://www.interstellarclouds.org}}. We create maps for each surveyed cloud  in Table \ref{tab:regions} using default parameters for the control field, a 2\farcm5 pixel size for the map and 2 pixel smoothing for the final map. Using iNICEST, the control field may be either a nearby field with little to no extinction, or it may be the science field for which we want a map. For the control field, the pipeline uses half of the control field area and selects the 50\% of stars with the lowest extinction. For three surveys---Hercules, Mon OB1, and Ophiuchus---we chose the science field as the control field as there was no more suitable field nearby. In testing, we found choosing the same field vs a separate control field had little effect on our results. This is because it mainly affects the extinction zero-point, which ultimately gets calibrated out, as shown in the next section.

\section{Calibrating Planck Dust Optical Depth}\label{sec:gamma}
\citet{2009ApJ...692...91G} showed that dust extinction provides the most reliable estimates of molecular gas column density in molecular clouds. The relationship between extinction and column density is fairly constant throughout the galaxy \citep[e.g.][]{2009ApJS..180..125R,2017MNRAS.471.3494Z}. However, where the column density is high enough, there are no stars visible with which to measure extinction. Dust emission provides an alternative for measuring dust column densities in the highest column density regions. However, to derive a mass from dust emission requires  knowledge of the dust temperature and the dust opacity law. Fits to the {\sl Planck} SEDs provide both the dust temperature and the spectral index of dust opacity law as described above. To derive column densities, the dust opacity, $\kappa_{d,353}$, is often assumed to be constant \citep[e.g.,][]{2020ApJ...890...42F,2021ApJ...912L..19P}; although, in reality, it may vary from cloud to cloud.

\begin{figure*}
    \centering
    \plotone{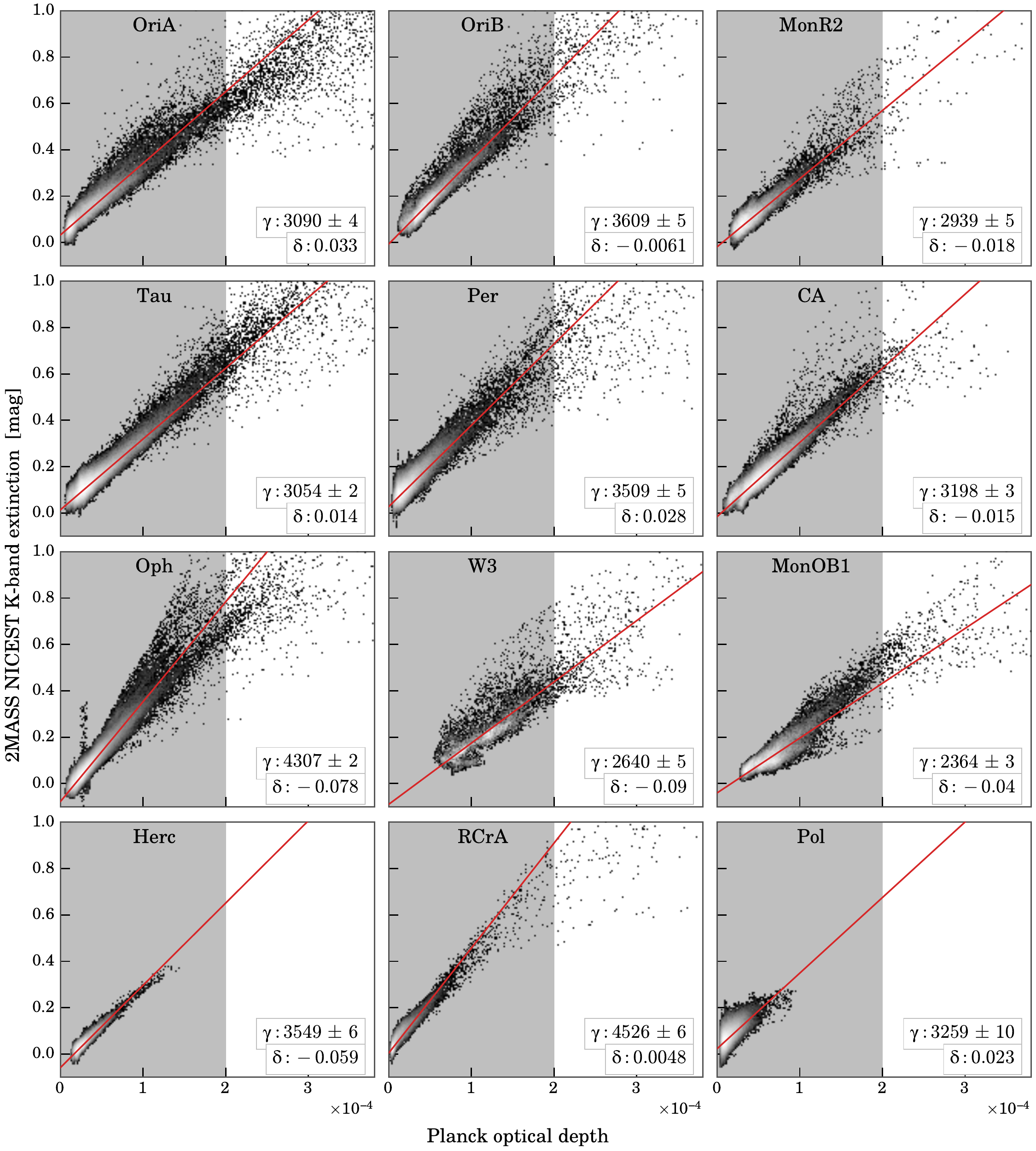}
    \caption{2MASS-based NICEST extinction (\A{K,2MASS}) plotted against Planck $\tau_{353}$. We restricted the fit to data within the gray border. The red line shows the best fit line from orthogonal distance regression, and we show the best-fit parameters in the lower right corner. Specifics on the fitting process can be found in \S\ref{sec:gamma}.}
    \label{fig:tauak}
\end{figure*}

\begin{figure}[ht]
    \centering
    \includegraphics[width=\columnwidth]{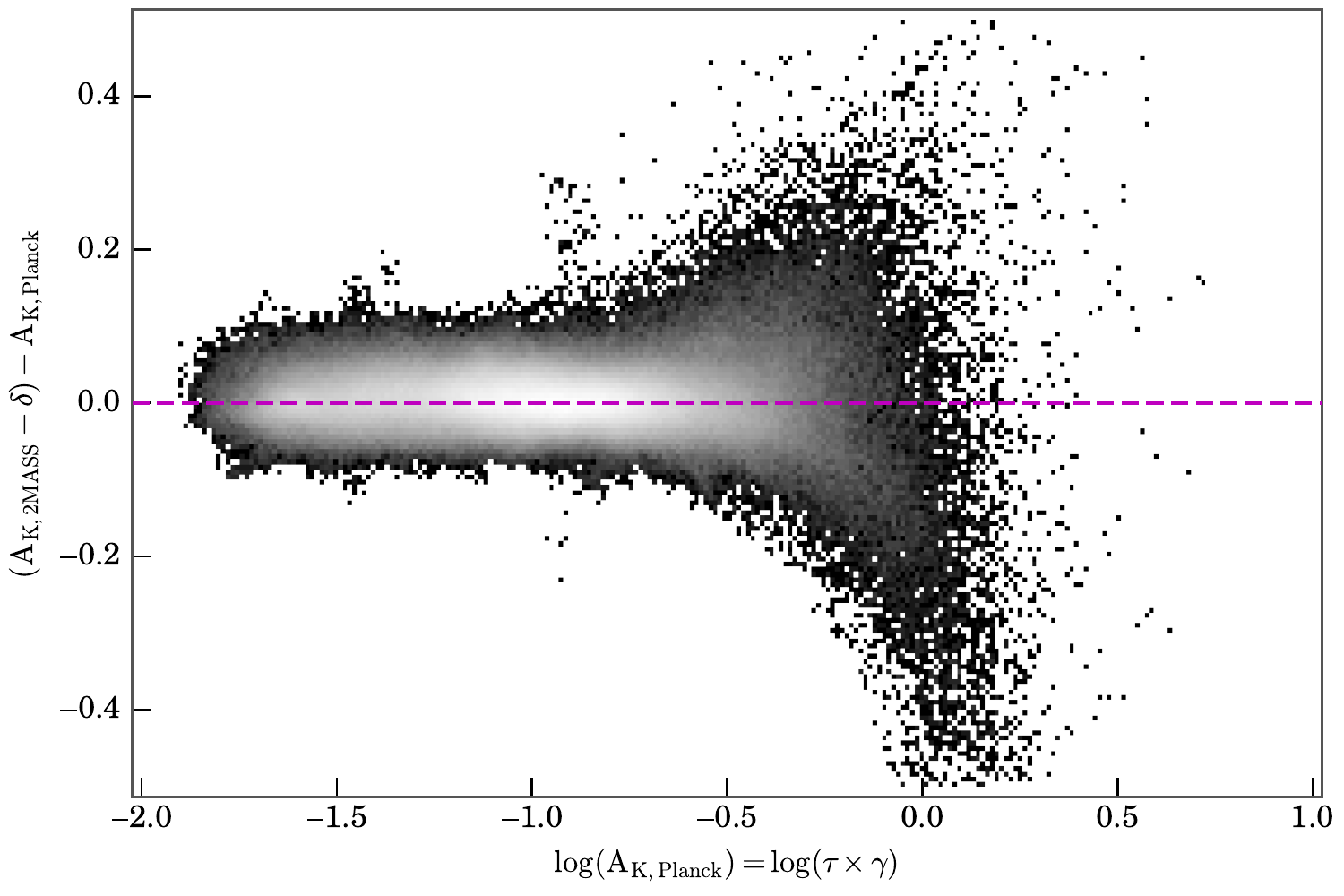}
    \caption{Difference of the NICEST extinction minus the calibrated \ak from Planck $\tau_{353}$. The size of the scatter is 0.027 mag, demonstrating the accuracy and consistency of this calibration. The large number of points trending downwards at $>-0.5$ lie where the extinction is too large to be measured using near-infrared extinction.}
    \label{fig:tauaksummary}
\end{figure}

We use the approach developed in \cite{2014A&A...566A..45L}, who calibrate the dust optical depth derived from \herschel dust emission with NICEST extinction to derive high-dynamic range maps of dust column density expressed as extinction. 
This approach combines the sensitivity of dust emission and the accuracy of dust extinction to create high-quality maps of column density. In Figure \ref{fig:tauak}, \planck dust optical depth (\taudust) is plotted against 2MASS dust extinction (\A{K, 2MASS}). \A{K, 2MASS} and \taudust show a clear linear relation. For $\taudust\approx 2\times10^{-4}$ the relationship becomes sublinear because the extinction map is not sensitive enough to probe such high column densities. To calibrate \taudust to extinction, we use orthogonal distance regression to fit a linear model, 
\begin{linenomath}
$$ A_{\rm K, 2MASS}\,=\,\gamma\,\taudust + \delta, $$
\end{linenomath}
where $\gamma$ is the conversion from dust optical depth to extinction and $\delta$ accounts for small calibration offsets in the NICEST extinction maps. Table \ref{tab:gammas} shows the results of the fits. The value of $\gamma$ spans a factor of 2 from $\sim2300-4500$ with a mean of $3350\pm 170$. The values of $\delta$ are quite small, with all $|\delta|<0.1$.  As \cite{2014A&A...566A..45L} discuss, using only the \planck and NICEST maps, we cannot determine if $\delta$ is an offset that needs to be applied to the \planck data. To determine this, we compare our 2MASS-based NICEST data with the 3D dust map from \citet{2019ApJ...887...93G}. The 3D dust map provides a measure of the color excess that is tied to absolute magnitudes of the stars, as opposed to being solely dependent on relative colors. If $\delta$ is an offset in the calibration of the NICEST maps, a similar offset should be present in a comparison with the 3D color excess $E_{\rm 3D}$. For each cloud, we fit a linear relation between \A{K,2MASS} and $E_{\rm 3D}$, $A_{\rm K, 2MASS}\,=\,R_{3D} E_{\rm 3D} + \delta^\prime$, deriving values for 3D color excess to {\it K}-band extinction ratio ($R_{3D}$) and the calibration offset ($\delta^\prime$). We find that  $\delta^\prime \approx 1.1 \delta$ (Pearson $r$ = 0.95). From this we conclude, $\delta$ represents a slight offset from the true intrinsic colors in the control fields used to derive the NICEST maps and is therefore not included in the calibration of the dust emission maps to extinction We can now express dust column densities in units of extinction as
\begin{linenomath}
\begin{equation*}
    A_{\rm K, \planck} \equiv \gamma \tau_{353}.
\end{equation*}
\end{linenomath}
A visual inspection of Figure \ref{fig:tauaksummary}, showing the combined residuals for all the clouds, shows that a linear fit is in general a good descriptor of the data. The residual scatter is $<0.03\, \magn$, and up to $\A{K, \planck}\sim0.3$ the scatter is well behaved and centered on zero. Beyond this, as explained previously, the 2MASS extinction systematically underestimates the extinction due to not sampling the highest extinction material. From this point, \ak will refer specifically to $\A{K, \planck}$.

For several clouds, this is the first measurement of $\gamma$; however, for several, measurements of $\gamma$ are reported in the literature through comparison with \herschel data.  \citet{2014A&A...566A..45L} reported $\gamma = 2637$ and $\gamma=3460$ in Orion A and B, respectively, while an updated value of $\gamma=3026$ was found for Orion A by using the PNICER method with deeper infrared photometry from the VISION survey \citep{2017A&A...601A.137M, 2018A&A...614A..65M}.  In Perseus, \cite{2016A&A...587A.106Z} find $\gamma=$3931. In California, \cite{2017A&A...606A.100L} measured $\gamma=$3593. \citet{2018A&A...620A..24H} measure a high value of $\gamma=$5256 in the Pipe, which is near the Ophiuchus region. 
For all the sources found in the literature, the values are within $15\%$ of the value we measure.

\subsection{Variation in the Dust Extinction -- Optical depth Ratio}\label{sec:gammavariation}
The value of $\gamma$ differs significantly in each cloud, and spans nearly a factor of 2.  
We expect this value to vary with dust properties. However, in addition to expected cloud-to-cloud variation, we identify a log-linear trend with distance of moderate significance, $\log\gamma = -0.18 \log{(d/\text{pc})} + 3.98$ (adjusted Pearson $r^2$=0.5, $p=0.002$). A sensitivity bias can result in an apparent trend with distance because setting a cloud at a farther distance imitates the effect a shallower survey depth has on measuring extinction. Quantitatively, the relationship between the change in survey depth ($\Delta m$) and distance ($d$) can be expressed as $\Delta m = -5\,\Delta\!\log{d}$. However, even over a small distance range, there is a considerable range in the value of $\gamma$.

To examine the effect of a sensitivity bias on our measurements of $\gamma$, we consider Orion A, which has a high sensitivity measurement of $\gamma$ from \cite{2018A&A...614A..65M}. The value of $\gamma=3090$ we derive in Orion A at 5\arcmin\ resolution is almost identical to the one ($\gamma=3026$) derived by \citet{2018A&A...614A..65M} who used extinction derived from deep VISION survey photometry \citep{2016A&A...587A.153M} to create 1\arcmin\ resolution extinction maps to compare with \herschel. They observed that in Orion A, a change in survey depth of $+4$ mag (the difference between the depth of the 2MASS and VISION surveys) resulted in an 15\% increase in $\gamma$ (the amount of the increase would vary depending on the true amount of dense material). Our good agreement with \citet{2018A&A...614A..65M} suggests that NICEST is able to do a better job recovering the contribution from unresolved extinction peaks at 5\arcmin\ than at 3\arcmin\ resolution, countering the effect of the sensitivity bias. This is because at lower resolution the average extinction at a location is lower and every pixel has a larger number of background stars, so the correction for unresolved substructure provided by NICEST is more robust than at high resolution. To test this finding, we generate an extinction map of Orion A with 2\farcm5 resolution and find that it misses as much as 1 mag of extinction near peaks when smoothed and compared to a map generated at 5\arcmin\ resolution. The coincidence of our $\gamma$ with that of \cite{2018A&A...614A..65M}, coupled with our test showing that at 5\arcmin\ we recover extinction better than at smaller resolutions, convinces us that for clouds nearer than Orion A our measurements of $\gamma$ should largely be unaffected by a sensitivity bias. If we assume we can scale the change in $\gamma$ with survey depth, as is seen in Orion, then for the more distant clouds (Mon R2, Mon OB1, W3), we estimate that a distance/sensitivity bias would result in a $~10\%$ increase in $\gamma$ at 2kpc. This is insufficient to account for the distance trend that motivated this investigation. It is apparent that survey sensitivity is not a dominant source of error in our determination of $\gamma$, and the origin of the distance trend is still uncertain. The observed trend with distance may still be related to an unrecognized bias in our extinction mapping method, or it may be an artifact due to our small sample size.

Examining the individual distributions in Fig. \ref{fig:tauak}, the clouds follow a linear relation reasonably well, considering the limitations of IR extinction mapping. 
However, toward higher extinction, the scatter has structure. 
Flattening is expected as extinction cannot sample the densest material. Meanwhile, we find that different branches in \A{K,2MASS} vs $\tau_{353}$, such as is seen in Ophiuchus and California, reflect changes in the dust $\beta$ derived by \planck. 
While there is no trend in $\beta$ with $\gamma$ that is consistent between the clouds,
within some clouds, pixels with $\beta<1.6$ have a different average $\gamma$ than pixels with $\beta>1.6$.
In the Orion A/B, California, and Ophiuchus clouds, a slightly higher $\gamma$ is associated with relatively warmer ($\tdust>20\ {\rm K}$) dust with $\beta < 1.6$.   
In clouds where \tdust spans more than a couple degrees Kelvin across the cloud, we see evidence of a $\beta$--$\tdust$ anticorrelation.

The effects of a possible sensitivity related bias and the variation of $\gamma$ with $\beta$ inside some clouds are both most noticeable at higher extinctions ($\ak\gtrsim 0.5$ mag). Nonetheless, a linear relation describes the bulk of the cloud's mass, which lies at low extinction, very well. All things considered, we adopt the custom $\gamma$'s we derive for each cloud in computing cloud masses.

\begin{deluxetable}{lccccc}
\tablecaption{Dust and CO Calibration}\label{tab:gammas}
\tabletypesize{\footnotesize}
\tablehead{\colhead{Name}  & \colhead{$\gamma$} & \colhead{$\delta$} & \colhead{\aco} & \colhead{Mass\tablenotemark{\footnotesize a}} &  \colhead{Radius}  \\  
 & & \colhead{(mag)} & \colhead{$\left( \frac{\msun\,  \pc^{-2}}{\Kkms} \right)$} & \colhead{($10^{3}\,\msun$)} &  \colhead{(pc)}
}
\startdata
Orion A        & 3090  & 0.033   & 4.00  & 85.2   & 23.0 \\
Orion B        & 3609  & -0.006  & 3.70  & 65.0   & 21.6 \\
Mon R2         & 2939  & -0.018  & 3.94  & 124.8  & 36.7 \\
Taurus          & 3054  & 0.014   & 3.88  & 19.2   & 13.1 \\
Perseus        & 3509  & 0.028   & 3.36  & 19.1   & 13.1 \\
California     & 3198  & -0.015  & 4.97  & 139.1  & 35.0 \\
Ophiuchus      & 4307  & -0.078  & 8.03  & 40.5   & 19.7 \\
W3             & 2640  & -0.089  & 5.63  & 241.8  & 39.3 \\
Mon OB1        & 2364  & -0.040  & 4.66  & 144.8  & 35.9 \\
Hercules       & 3549  & -0.059  & 4.01  & 4.3    & 7.3 \\
Corona         & 4526  & 0.005   & 4.99  & 1.2    & 3.2 \\
Polaris  & 3259  & 0.023   & 2.82  & 2.3    & 5.8
\enddata
\tablenotetext{a}{Mass and radius derived { from dust extinction} for $\ak>0.1\ \magn$}
\tablecomments{Derived values of $\gamma$ ($\A{K,2MASS}/\tau_{353}$ conversion factor), $\delta$ (2MASS extinction calibration offset), and \aco (CO luminosity to total gas mass conversion factor) for each cloud. Errors are not given for $\gamma$ and $\delta$ as linear fitting drastically underestimates the real error for this data set. The error in \aco $\lesssim1\%$. }
\end{deluxetable}

\subsection{Relation of the Dust Extinction--Optical Depth Ratio to the Dust Opacity}

The calibration parameter $\gamma$ is related to the dust opacity, $\kappa_{\rm d,\nu}$. From the definition of optical depth, $\dd\tau_\nu = \kappa_\nu\rho\,\dd z$,
\begin{linenomath}
\begin{equation}\label{eqn:opacitydef}
\kappa_{\rm d,353} = \tau_{353}\frac{ R_{\rm dg}}{\N{H}\,\mu m_p}, 
\end{equation}
\end{linenomath}
where \N{H}=\nhtotal, $m_p$ is the mass of a proton, $\mu$ is the mean molecular weight, and $R_{\rm dg}$ is the dust-to-gas mass (or mass density) ratio. \N{H} can be derived from extinction, $\N{H} = \beta_K \ak$, where $\beta_K \equiv \N{H}/{\ak} = 1.67\E{22} \dgunit$ \citep{1979ARA&A..17...73S,1978ApJ...224..132B}. This common value of $\beta_K$ assumes $\ak/\av = 0.112$ and $R_{\rm V} = 3.1$. Since \planck extinction $A_{K,\planck} = \gamma\, \tau_{353}$, 
\begin{linenomath}
\begin{equation}\label{equ:opacitygamma}
\kappa_{d,353} = \frac{R_{\rm dg}}{\mu m_p \beta_K \gamma}.
\end{equation}
\end{linenomath}
For our average $\avg\gamma \approx 3300$, $\mu = 1.37$ (for standard 64\% H, 36\% He ISM composition; e.g., \citealt{2010ApJ...723.1019H}) , and assuming $R_{\rm dg}\approx 100$, we find that on average $\kappa_{d,353} = 0.8\ {\rm cm^{2}\, g^{-1}}$. This is about $\sim$2-3$\times$ larger than expected for the diffuse ISM in the Milky Way \citep{2003ARA&A..41..241D}, but is just less than $\kappa = 1.1$ for dust with a thin-ice mantle in protostellar cores from \citet{1994A&A...291..943O}. Our value lies in between $\kappa$ for the diffuse ISM and GMC cores. This seems appropriate as our survey covers $\sim$parsec scales, which straddles the two regimes. Given $\kappa_{d,353} \sim R_{\rm dg}\gamma^{-1}$, larger values of $\gamma$ imply lower opacities ($\kappa$).

\subsection{Total gas surface density from extinction}

 We derive the total gas surface or column density, $\Sigma_{gas}$, including \HI, \htwo, and  atomic He, from extinction using the equation,
\begin{linenomath}
\begin{equation}
    \Sigma_{\rm gas} = \mu m_p \beta_K\, \ak,
\end{equation}
\end{linenomath}
where $\mu = 1.37$, $m_p$ is the proton mass, and $\beta_K = \N{H}/\ak = 1.67\E{22} \dgunit$, the same as above. The $\ak:\Sigma_{\rm gas}$ conversion factor $\sigcon \equiv \Sigma_{\rm gas}/\ak = \mu m_p \beta_K = 183.2\, \msunpc \magn^{-1}$. The cloud's total mass is found by integrating the column density over a area ($S$) of the cloud, 
\begin{linenomath}
\begin{equation}\label{eqn:mgas}
M(\ak)  = \int \Sigma_{\rm gas}\, \dd S  = \sigcon \sum \ak\, \Delta s.
\end{equation}
\end{linenomath}

where $S$ is the area being integrated over. In the \texttt{CAR} projection used in the DHT survey, the pixel area is $\Delta s  = 2 d^2 \Delta\glon \sin(\Delta\glat/2) \cos(\glat)$, where $d$ is the object distance (or $d$=1 for angular area) and $(\glon, \glat)$ are defined at the center of the pixel. This change is important on large scale maps.

\section{Calibrating CO as a Mass Tracer}\label{sec:cointro}
We can now investigate the effectiveness of CO as a tracer of the total gaseous mass  by comparing the calibrated extinction maps we derived from dust emission with the CO maps from the \citetalias{2001ApJ...547..792D} survey.

\subsection{The relationship between cloud mass and CO luminosity}\label{sec:cloudscale}

\begin{figure}
    \centering
    \includegraphics[width=\columnwidth]{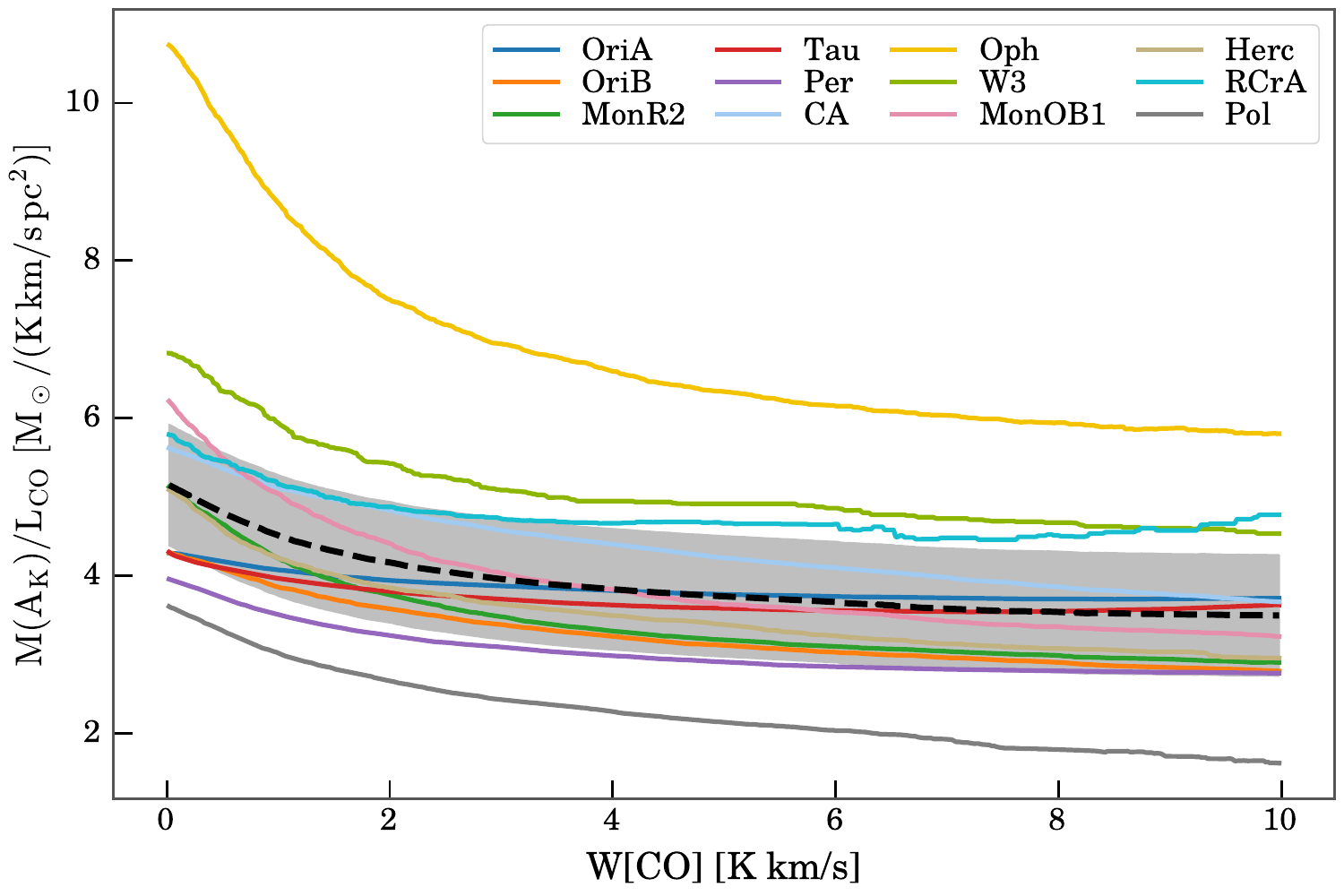}
    \caption{$\alpha_{\rm CO}$ plotted as a function of \wco-defined cloud boundary (\W0). The colored lines are the individual clouds and the black trace and the shaded gray region shows the average \aco and the standard deviation (excluding the extremes Ophiuchus and Polaris).}
    \label{fig:alphaco}
\end{figure}

The conversion factor, \aco,  used to transform a measurement of CO luminosity to the total gaseous  mass of an entire cloud is defined as:
\begin{linenomath}
$$ \aco \equiv  \frac{\M{gas}}{L_{\rm CO}}. $$
\end{linenomath}

Here $M_{\rm gas}$ is the total gaseous mass. We assume that the total gaseous mass of a cloud, $M_{\rm gas}$,  is given by our dust-derived mass defined in Equation \eqref{eqn:mgas} (i.e., $M_{\rm gas} = M(\ak)$). $L_{\rm CO}$ is the total CO luminosity measured in units of $(\Kkms\ \pc^{2})$ and is obtained from integrating \wco, the CO integrated intensity, over the CO emitting area of the cloud, i.e., 
$L_{\rm CO} = \int\!\wco\,\dd S $. 
The CO luminosity of a cloud is defined relative to a minimum value or boundary, \W0, of the map of integrated intensity.   
In Figure \ref{fig:alphaco}, we show the value of \aco as a function of varying values of \W0 for each cloud in our sample. We also plot the mean (black trace) and error (standard deviation, shaded gray region) of \aco at each value of \W0. 
The rise seen at low boundary values is likely due to the CO abundance decreasing toward the edge of the clouds, itself likely due to photoionization of the CO molecules there. 
We define our preferred value of \aco using the $\wco>1.5\,\Kkms$ boundary. 
This corresponds to the average $5\sigma$ detection limit for our sample of clouds and defines where we consider CO as being significantly {\it detected}. Moreover, above this level the value of the derived \aco in individual clouds is relatively stable.
The value of \aco for each cloud can be found in Table \ref{tab:gammas}.
In determining the average we removed the two extreme values coming from Ophiuchus and Polaris, this gives a value $\aco=4.31\pm0.67\ \alphaunit$ for the $J=$1--0 transition which is in close agreement with the common value from \citet{2013ARA&A..51..207B}.  If we don't exclude the extreme values, the average increases to $4.5\pm 1.3\ \alphaunit$. We adopt $\aco=4.31\pm0.67\ \alphaunit$ as our value for \aco throughout the rest of the paper.

Multiplying \aco by the measured total luminosity of CO above our detection limit gives the mass of the CO emitting area. We note that because of the existence of CO dark molecular gas, the mass of the CO emitting area derived using \aco is not necessarily the mass of all the molecular gas in a molecular cloud \citep[e.g.,][]{2010ApJ...716.1191W}. However, dust emission and absorption measurements can sample the entire extent of a molecular cloud and produce more accurate and robust cloud masses. Here, we are primarily concerned with calibrating an accurate as possible mass for CO emitting gas in GMCs. This is essential for using CO to trace gas masses in more distant regions in the Milky Way and particularly in external galaxies, where observations of dust on GMC scales are not typically possible. 

\subsection{The X-factor}\label{sec:xfactor}

We note here that the more traditional CO conversion factor, the {\it X}-factor (i.e.,$ \xco \equiv {\frac{\N\htwo}{\wco}} $)  is equivalent to \aco, apart from a scaling constant. Specifically $\aco = 2\mu m_p \xco = 9534 \mu m_p$\xco (in units of $\msunpc [\Kkms]^{-1}$) and thus \xco has also been frequently used to provide measures of molecular cloud masses within the Galaxy and beyond. Our preferred value of \aco corresponds to \xco = $1.99 \pm 0.31 \E{20} $, essentially the canonically accepted value \citep[e.g.,][]{2013ARA&A..51..207B}. 

Recently, \citet{2020ApJ...898....3L} (hereafter, LD20) compared CO and extinction derived masses for hundreds of GMCs within a few kiloparsecs of the Sun. These masses were extracted from independent CO \citep{2016ApJ...822...52R,2017ApJ...834...57M} and extinction \citep{2020MNRAS.493..351C} based cloud catalogs. From this exercise, \citetalias{2020ApJ...898....3L}  derived \xco $= 3.6 \pm 0.3 \times 10^{20} $---a significantly higher value than derived here.  The difference between their derived value and the canonical value is likely the result of the fact that the average surface densities of GMCs  derived from both CO cloud catalogs using the canonical value of \xco ($\Sigma_{\rm GMC}$ = 26.8 $\pm$ 0.7 and 20.3 $\pm$ 1.3 \msunpc) are significantly smaller than both that  derived from early extinction studies of local clouds ($\Sigma_{\rm GMC}$ = 41.2 $\pm$ 1.6 \msunpc)  (\citealt{2010ApJ...724..687L}; \citetalias{2010A&A...519L...7L}) and that ($\Sigma_{\rm GMC} =51.5 \pm 1.1\ \msunpc$) derived from the larger and more recent \citet{2020MNRAS.493..351C} infrared survey. Moreover, in \S\ref{sec:constsurf}, we find a CO derived average surface density  of the clouds in our sample ($\Sigma_{\rm GMC} =$ 37\msunpc) to be closer in agreement with the infrared studies discussed above. As discussed by \citetalias{2020ApJ...898....3L} and demonstrated later in the present paper, the average column density of a cloud depends on the boundary level used to define the cloud. We postulate that the differences in the X-factor values derived from comparison of  the two CO-based cloud catalogs \citep[i.e.,][]{2016ApJ...822...52R,2017ApJ...834...57M} with the infrared-based catalog \citep{2020MNRAS.493..351C}  and with the {\it X}-value derived here have their origin in the use of different and varying cloud boundaries to define the clouds in the respective studies. Further discussion of this point is deferred until later in this paper.

\section{Molecular Cloud Structure: The N-PDF}\label{sec:mass_size0}

\subsection{The Extinction PDF}\label{sec:npdf}
\begin{figure*}[t]
    \centering
    \includegraphics[width=\textwidth]{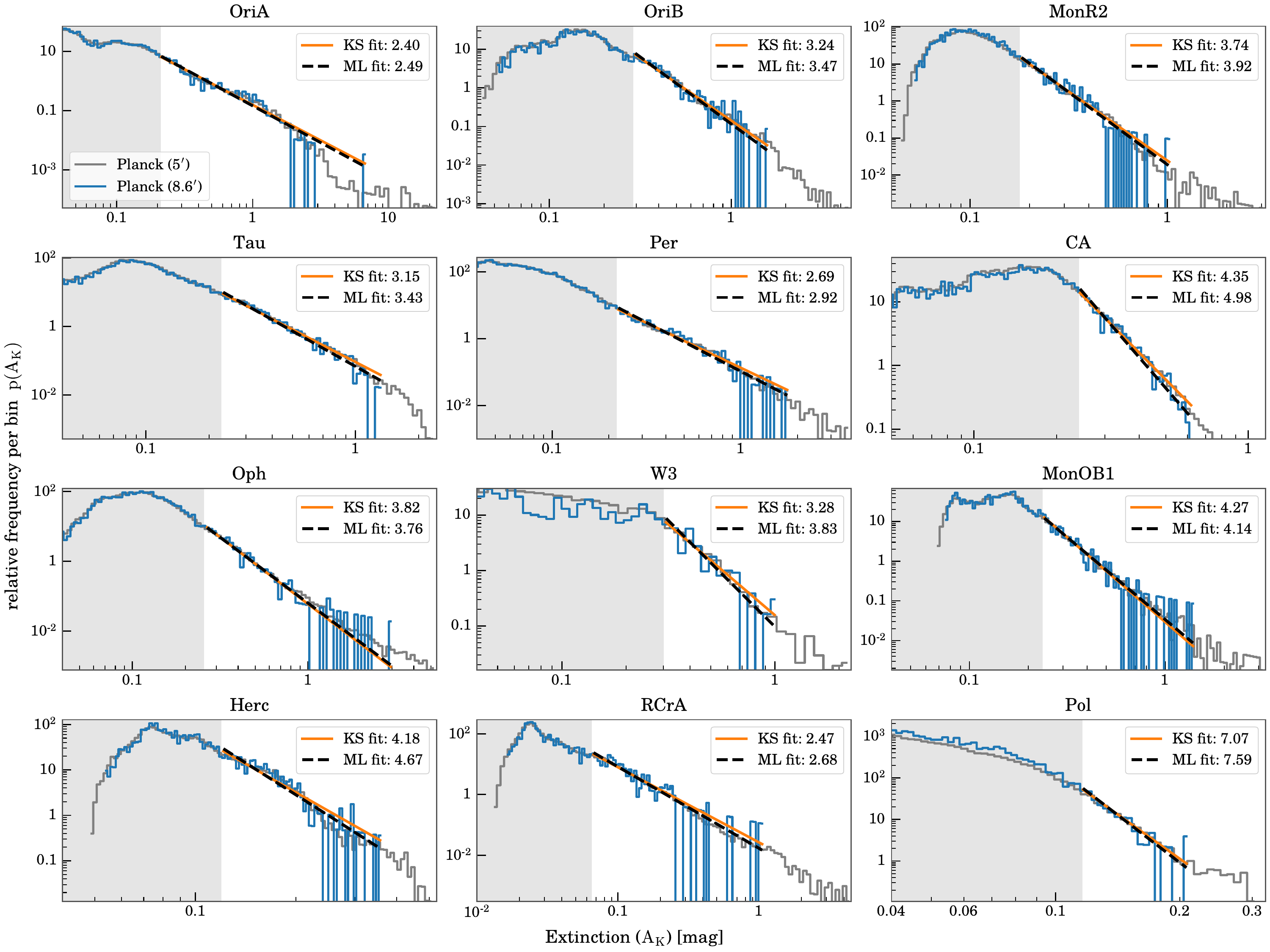}
    \caption{Extinction PDFs for each cloud at 5\arcmin (gray) and 8\farcm6 (blue) resolution. The gray shading indicates where we are incompletely sampling the PDF, i.e., where we are below the lowest closed extinction contour. The black dashed line is the ML fit, and the solid orange line is the result of a fit of a truncated-power law. The fits { and closed contour are derived using} the 8\farcm6 resolution \planck data. { The PDFs are normalized so that the integral above the closed contour equals one.}}
    \label{fig:pdfs}
\end{figure*}

A fundamental metric frequently used to describe the structure of a GMC is the column density probability density function or N-PDF. The N-PDF of a GMC is equivalent to the PDF of cloud extinctions.
Above their lowest closed extinction (column density) contour, molecular clouds have been shown to have power-law  N-PDFs \citep{2015A&A...576L...1L}. In Figure \ref{fig:pdfs}, we show the extinction derived PDFs for the 12 clouds in our sample for maps at both the native \planck resolution (5\arcmin, gray) and the resolution of the CO survey (8\farcm6, blue). The following analysis is based on the 8\farcm6 maps to facilitate comparison with CO. 
As shown by \citet{2017A&A...606L...2A}, extinction  PDFs are only complete for extinctions above the lowest closed contour in an extinction map and in Figure \ref{fig:pdfs} the gray shaded regions indicate the range of extinctions where the PDFs are incomplete. Here, we define the lowest closed contour level to be the 99\% completeness limit---i.e., the lowest contour for which 99\% of the enclosed area is contained within the bounding box for the cloud. The PDFs are normalized so that the integral above the closed contour equals one. Above the respective completeness limits we fitted power laws to the data using two methods --- the maximum-likelihood (ML) method described in \citet{2009SIAMR..51..661C} and implemented in \texttt{powerlaw} Python package \citep{2014PLoSO...985777A}, and by minimizing the Kolmogorov–-Smirnov (K-S) test statistic between the data and a truncated power-law distribution. The truncation ($A_{\rm K,max}$) was set to the peak extinction in the cloud. The fits are plotted in the figure and the results are given in Table \ref{tab:pdf}.  The PDFs are reasonably well described by single power-law functions, with the derived power-law indices from the two fitting methods generally agreeing to 10\% or better. 
We adopt the ML results as our value for the power-law slope. Fits derived with the native resolution extinction maps produce slopes that, on average, agree to 10\%.
The indices or slopes of the fitted power laws range from $\sim 2.5 - 8$. However, ignoring the Polaris cloud, which seems to be an outlier, the range is from $\sim$ 2.5--5. The mean slope or index of the distributions is $\langle n\rangle = 3.66 \pm 0.74$, excluding Polaris. \citet{2017A&A...606L...2A} found $n\approx 4$ for the inner regions of Polaris using \herschel data. The difference is due to the reduced resolution of the study, as the closed contour,{ above which the slope is determined, has a similar value and} covers a similar area as \citet{2017A&A...606L...2A}. { The low resolution combined with the small area within the closed contour results in Polaris containing very few pixels ($\sim 10^2$) relative to other sources ($\sim 10^3$). If we use the 5\arcmin\ resolution extinction map and 0.2 mag as the closed contour, we find $n\sim 3.7$ in Polaris, in line with \citet{2017A&A...606L...2A}.}

\begin{figure}[t]
    \centering
    \includegraphics[width=0.5\textwidth]{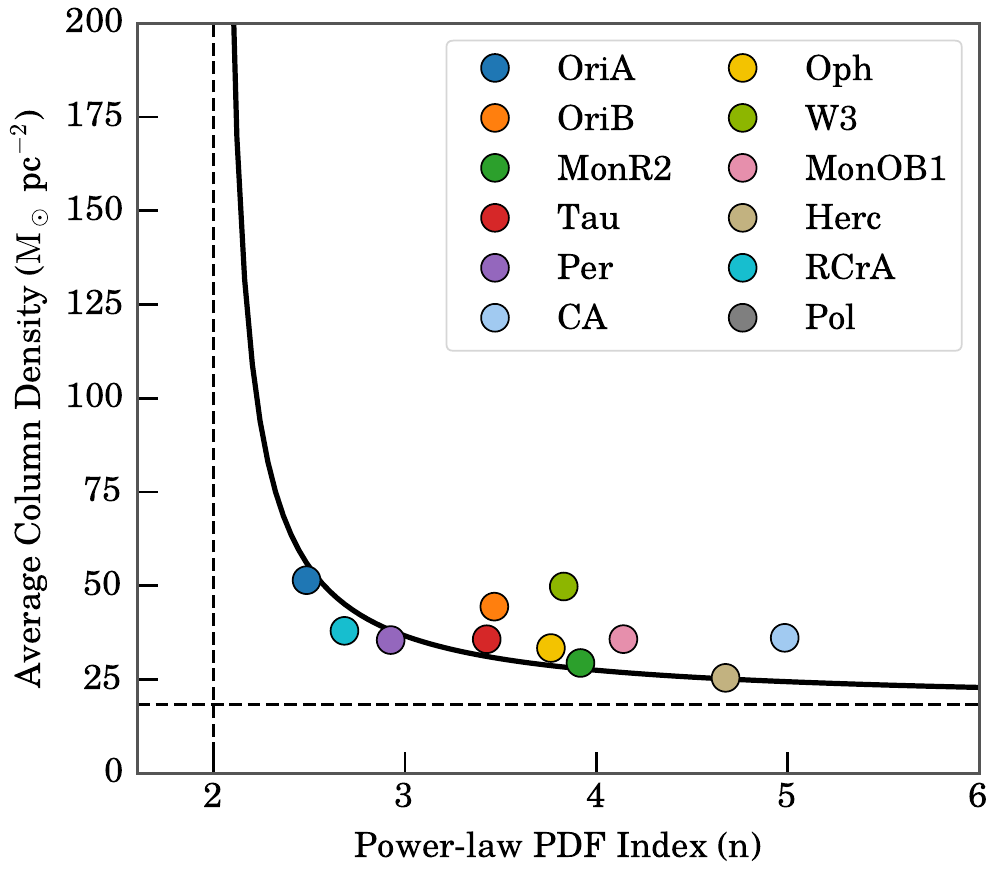}
    \caption{The predicted average column density for a GMC with power-law PDF index $n$, measured above the 0.1 mag extinction contour. The black dashed lines indicate {(vertical)} an index of $n=2$ and {(horizontal)} the average column density corresponding to a {uniform extinction field} of $\ak=0.1$ mag $\sim$ 18.3 \msunpc. The points correspond to the \avg\ak (above 0.1 mag) and $n$ for each cloud. }
    \label{fig:surfind}
\end{figure}

Our results  confirm the findings of \cite{2015A&A...576L...1L}, indicating that, down to the completeness limits of the observations, molecular clouds exhibit power-law PDFs and furthermore that there is no evidence for the log-normal shapes generally predicted by turbulent simulations \citep[e.g.,][]{1997ApJ...474..730P,1998PhRvE..58.4501P,2017ApJ...843...92B,2001ApJ...546..980O,2003ApJ...592..975L}. The completeness limits of the observations reported here generally range from  0.1--0.3 mag of ({\it K}-band) extinction. These extinctions are above that of the atomic-molecular transition boundary for molecular clouds ($\approx$ 0.05 mag at 2.2 $\mu$m, \citet{2017ApJ...843...92B, 2016ApJ...829..102I})
confirming that the clouds we are observing are fully molecular, and consequently, that the internal structure of the molecular gas in GMCs is clearly characterized by power-law PDFs. In this context it is interesting to note that the one cloud (RCrA) where the completeness limit is measured to be $\sim$ 0.06 mag, close to the expected transition between atomic dominated and molecular dominated gas, the PDF is still power law in form with no suggestion of a log-normal behavior.

The level of star formation activity in GMCs has been empirically linked to cloud structure, specifically to the nature of the distribution of column densities within a cloud \citep[][]{2009A&A...508L..35K,2010ApJ...724..687L,2013ApJ...778..133L}. The fact that the PDFs of GMCs are power laws also provides some interesting insights regarding their measured properties.  In particular, we can determine  the average column or surface density of a cloud with knowledge of its PDF.  The average surface density is an important cloud property that is connected to two {important} scaling laws for star formation, the Kennicutt--Schmidt relation and the mass--size relation. For a power-law PDF with an index  $n>1$, the PDF is given by
\begin{linenomath}
\begin{equation}
p(\ak) = {\frac{n-1}{\akmin}} \left(\frac{\ak}{\akmin}\right)^{-n} \label{eqn:plpdf}
\end{equation}
\end{linenomath}
where \akmin is the extinction threshold above which the PDF is a power law. Typically, this corresponds to the completeness limit of the PDF. Following the analysis of \citet{2012MNRAS.427.2562B} the mean extinction for this distribution can be meaningfully calculated for the case where $n > 2$ as
\begin{linenomath}
\begin{equation}
    \langle\ak\rangle = \frac{\int_{\A0}^\infty\ak\, p(\ak)\ \dd\ak}{\int^\infty_{\A0} p(\ak)\ \dd\ak} = {\frac{n-1}{n-2}} \A0 \label{eqn:avgext}
\end{equation}
\end{linenomath}
Thus, the mean extinction (column density) of a cloud with a power-law PDF is determined by only two parameters---the power-law index, $n$, and the minimum or threshold extinction, $A_0\ge\akmin$. This threshold extinction (\A0) is typically set at the adopted cloud boundary in an extinction map, which optimally is given by the completeness limit. The mean extinction is equivalent to the mean surface density apart from a constant, $\avg{\Sigma_{\rm gas}} = \sigcon \ak$. For a fixed boundary, changes in $n$ will result in different values for the average extinction. In Figure \ref{fig:surfind}, we plot the cloud mean extinction or column density for a range in power-law index $n$. We overplot the average column density for each cloud in our sample. As $n$ approaches 2, the relation becomes divergent and small changes in $n$ can lead to large changes in the derived column density. This is because for shallower slopes, a larger fraction of the cloud mass is at high extinction. If $n$ takes on either a narrow range of values or is $\gtrsim\!2.5$, then the value of $\frac{n-1}{n-2}$ will be relatively constant. For a fixed value of $n$, the average extinction measured for a cloud will vary directly with the value of $A_0$. This dependence of measured cloud surface densities on the adopted cloud boundary, $A_0$, has already been indicated in a set of earlier observations of local clouds where \citetalias{2010A&A...519L...7L} found that $\avg\ak/A_0 \approx 2$, which in turn implied that  $n \approx 3$, a value consistent with the mean index ($\avg n = 3.17 \pm 0.46$) subsequently found to characterize the PDFs of many of the same clouds \citep{2015A&A...576L...1L}. In our sample, we find $\avg n = 3.66 \pm 0.74$ and $\avg\ak = \avg{\frac{n - 1}{n - 2}} \A0$ $ = (1.8 \pm 0.5) \A0$, where the error is the standard deviation.

The average column or surface density can be an important metric for characterizing cloud properties in comparative studies of GMCs across differing scales and in differing environments. Given that GMC PDFs are power laws, care must be taken in establishing consistent cloud boundaries before performing meaningful comparisons of surface densities between cloud populations both within and between galaxies (e.g., \citetalias{2020ApJ...898....3L}).

\begin{deluxetable}{lcccc}
\tablecaption{Power-law N-PDF fits}
\label{tab:pdf}
\tablehead{\colhead{Name} & \colhead{$n$ (ML)\tablenotemark{\footnotesize a}} & \colhead{$n$ (KS)\tablenotemark{\footnotesize b}}&  \colhead{$A_{\rm K, min}$\tablenotemark{\footnotesize c}} & \colhead{$A_{\rm K, max}$\tablenotemark{\footnotesize d}} }
\startdata
Orion A &  2.49 &  2.40 &  0.21 &  6.85 \\
Orion N &  3.47 &  3.24 &  0.29 &  1.60 \\
Mon R2 &  3.92 &  3.74 &  0.18 &  1.04 \\
Taurus &  3.43 &  3.15 &  0.23 &  1.34 \\
Perseus &  2.92 &  2.69 &  0.22 &  1.80 \\
California&  4.98 &  4.35 &  0.24 &  0.63 \\
Ophiuchus &  3.76 &  3.82 &  0.25 &  3.02 \\
W3 &  3.83 &  3.28 &  0.30 &  1.04 \\
Mon OB1 &  4.14 &  4.27 &  0.24 &  1.41 \\
Hercules &  4.67 &  4.18 &  0.12 &  0.36 \\
R Corona A &  2.68 &  2.47 &  0.06 &  1.08 \\
Polaris &  7.59 &  7.07 &  0.12 &  0.22
\enddata
\tablenotetext{a}{The power-law PDF slope determined using the ML method from \citet{2009SIAMR..51..661C}. This is the preferred value used in the text}
\tablenotetext{b}{The power-law PDF slope determined using K-S test minimization for a power-law distribution truncated to \akmax}
\tablenotetext{c}{\akmin is the lowest closed extinction contour}
\tablenotetext{d}{\akmax is the peak value of the extinction map within the cloud boundary}
\end{deluxetable}

\subsection{The CO PDF}\label{sec:copdf}
\begin{figure*}[t]
    \centering
    \includegraphics[width=\textwidth]{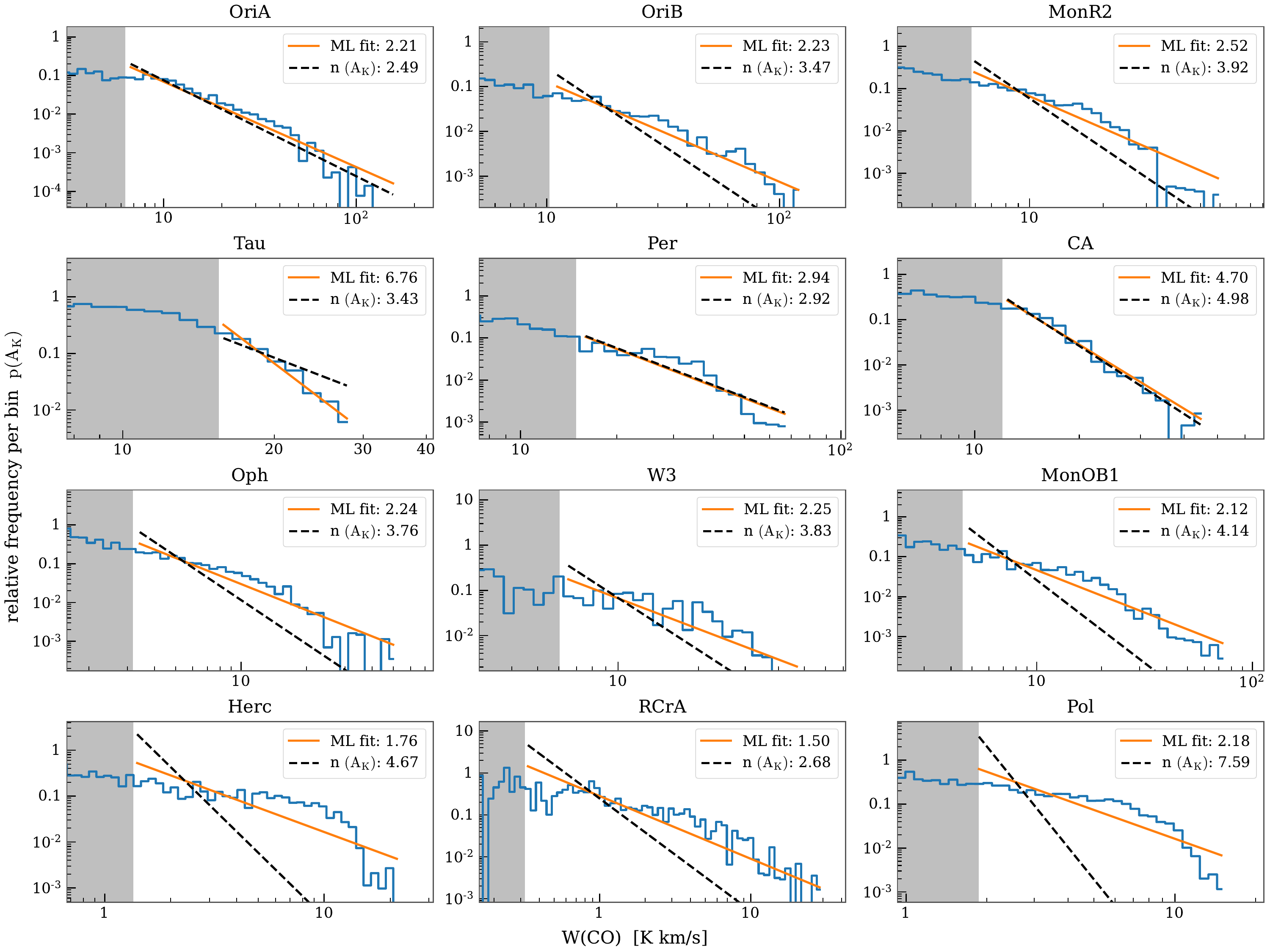}
    \caption{\wco probability density functions for each cloud.The gray shading indicates where we are incompletely sampling the PDF, i.e., where we are below the lowest closed \wco contour. The black dashed line is the fit from extinction, and in orange is the result of a maximum likelihood (power-law) fit to the CO emission data. { The PDFs are normalized so that the integral above the closed contour equals one.}}
    \label{fig:copdfs}
\end{figure*}

Because $^{12}$CO emission is generally optically thick and becomes both saturated at modest  extinctions and depleted at high extinctions, \wco is a poor tracer of column density on sub-cloud scales. Consequently, the CO PDF of a GMC is not equivalent to the N-PDF of the cloud. Nonetheless, it is instructive to consider how the CO PDFs compare to the extinction PDFs in GMCs.  Figure \ref{fig:copdfs} shows the PDFs of \wco for the clouds in our sample. 
The CO PDFs are relatively flat below their respective completeness limits and then rapidly decline toward high values of \wco.
Power laws fit to these data are displayed in the figure along with the values of the slopes.  Unlike the extinction PDFs in Figure \ref{fig:pdfs}, the CO PDFs are not particularly  well described by single power-law functions. Indeed, for most sources, the slopes of the extinction and CO PDFs are significantly different in value. The discrepancies between the extinction and CO PDFs are likely due to the fact that CO does not effectively trace column density on sub-cloud scales. This supposition finds some support in Appendix \ref{app:xco}. There we show that the {\it X}-factor exhibits significant variations on sub-cloud scales, and that the extent of these variations change from cloud to cloud. Clearly our observations demonstrate the $^{12}$CO PDFs of molecular clouds provide poor metrics for describing  cloud structure and, not surprisingly, are clearly inferior to extinction PDFs in this regard.

\section{The Mass -- Size scaling relation}\label{sec:mass_size1}
\subsection{Constant Surface Density GMCs}\label{sec:constsurf}
Some 40 years ago, \citet{1981MNRAS.194..809L} identified an interesting empirical scaling relation between the masses and sizes of nearby molecular clouds. Specifically, \citeauthor{1981MNRAS.194..809L} found $\M{GMC} \propto \R{GMC}^{1.9}$ suggesting that the mean surface densities of GMCs ($\Sigma_{\rm GMC}$) were constant, that is, $\M{GMC}/\R{GMC}^{1.9}  \approx \Sigma_{\rm GMC} =$ constant. \citeauthor{1981MNRAS.194..809L} did not estimate the value of constant column density implied by this scaling relation. However, a few years later, \citet{1987ApJ...319..730S} derived a value of $\avg{\Sigma_{\rm GMC}} =$ 170 \msunpc from analysis of a CO survey of GMCs in the inner Galaxy. Over the next few decades, observations of CO found GMC surface densities to range between 2 and 200 \msunpc \citep{2015ARA&A..53..583H}. These results were obviously in tension with the idea of constant column density GMCs, although few measurements of the mass--size relation were actually reported in the literature during this time. However, \citetalias{2010A&A...519L...7L} revived interest in the mass--size relation when, using observations of infrared extinction, they showed conclusively that local clouds followed a very tight mass--size relation with $\M{GMC} \sim \R{GMC}^2$ resulting in an average column density for GMCs constant to within $\sim$ 10$\%$. Specifically, they found $\avg{\Sigma_{\rm GMC}} = 41.2 \pm 5$ \msunpc for their local cloud sample. This value for local GMCs was in addition  much lower than that implied by the work of \citet{1987ApJ...319..730S}. They also showed that the derived value of $\avg{\Sigma_{\rm GMC}}$ depended on the cloud boundary adopted for the measurement. The implications of such a tight distribution of GMC column densities is potentially significant for understanding both molecular clouds and the star formation process within them. Consider, for example, that a population of GMCs characterized by a constant column density cannot obey the Kennicutt--Schmidt law \citep[e.g.,][]{2013ApJ...778..133L}

\begin{figure*}[t]
    \centering
    \includegraphics[width=\textwidth]{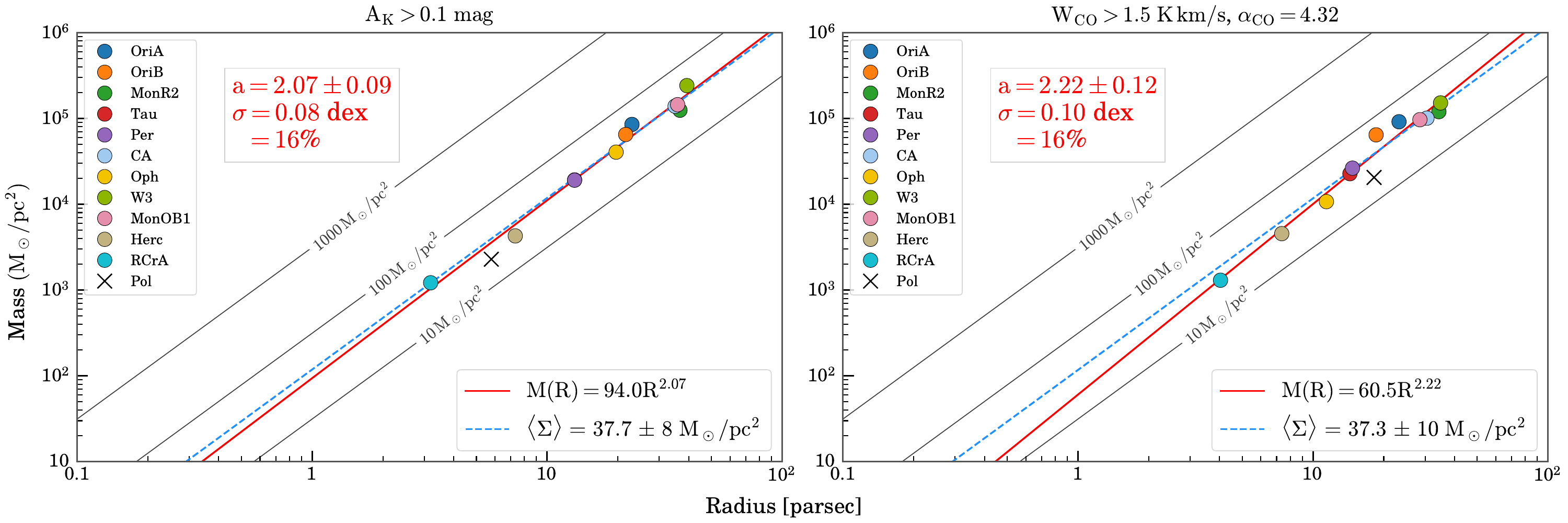}
    \caption{The mass--size relation derived from dust extinction with \ak$>$0.1 mag (left) and CO emission with \wco$>$1.5 \Kkms\  (right). The labeled gray lines are at constant surface density. { Polaris is shown as a black "$\times$" as it is not included in the fit or derived values shown.}}
    \label{fig:massradius}
\end{figure*}

Recently, \citetalias{2020ApJ...898....3L} re-examined existing CO and infrared extinction observations of the Milky Way in an attempt to investigate the discrepancy between the idea of a constant column density for GMCs, indicated by the mass--size relation,  with the wide range of cloud surface densities derived from existing CO measurements of Galactic clouds. They constructed the mass--size relation for GMCs across the entire Milky Way disk using data in existing catalogs of molecular clouds. They confirmed that the mass--size relation for Milky Way GMCs is best described by a power law with an index $\sim$ 2 in both the CO and infrared data. However, the scatter in the CO derived relations was uncomfortably large compared to that of the infrared data. They determined that the additional scatter seen in the CO relations was due to a systematic variation of $\Sigma_{\rm GMC}$ with galactocentric radius that was unobservable in the infrared measurements, which were confined to a local and much smaller volume of the Galaxy than the CO observations. After correcting the CO data for the Galactic metallicity gradient, they found that, outside the molecular ring, the GMCs in both the inner (i.e., $\R{gal} < 4$ kpc) and outer (i.e., $\R{gal} > 7$ kpc) Galaxy could be characterized by roughly the same constant surface density of 35 \msunpc. Within the molecular ring ($4 \leq \R{gal} \leq 7$ kpc) the average GMC surface density was measured to be higher (82 \msunpc). They argued that the measurements in the Galactic Ring were likely contaminated and biased to higher values by severe cloud overlap and blending in that region of the Galaxy.

In Figure \ref{fig:massradius} we show the mass--size relations respectively derived from the infrared extinction and CO data for the clouds in our sample. The cloud masses were calculated by simply integrating the surface densities over the cloud area,
\begin{linenomath}
\begin{equation*}
    M_{\rm GMC} = \int\limits_{\Sigma_{\rm gas}>\Sigma_0}\!\!\!\!\! \Sigma_{\rm gas}\, \dd S,
\end{equation*}
\end{linenomath}
where $\Sigma_{\rm gas}$ is either \sigcon\ak or \aco\wco for extinction and CO, respectively, and $\Sigma_0$ is the adopted cloud boundary. There are various estimators for the {\it size} of a molecular cloud. \citet{1981MNRAS.194..809L} used the length of the longest axis, while others have suggested using the geometric mean of the major and minor axes \citep{2006PASP..118..590R,2008ApJ...679.1338R}. We simply use the radius $R = \sqrt{S/\pi}$ for the size, where $S$ is the area of the cloud. 
The masses and sizes of the GMCs in our sample were calculated assuming the same boundary for all clouds{, using the 8\farcm6 resolution maps. A reasonable definition for a molecular cloud is that the cloud must predominantly consist of molecular gas. We therefore select as a boundary $\ak \geq 0.1\magn$\ which is the transition between C and CO gas (i.e., where CO is protected from photodissociation by self-sheilding) and is close to the \HI--\htwo cloud interface ($\sim 0.05 \magn$). We do not go down to 0.05 mag because at that level  spatially extended, unrelated foreground and background emission begin to dominate our measurements. Using this boundary also  facilitates comparison with \citetalias{2010A&A...519L...7L}. } 
For the CO mass--size relation, we adopt the $\wco \geq 1.5\ \Kkms$ for the boundary defining the cloud, as discussed in \S\ref{sec:cloudscale}. 
Neither boundary is a closed contour for most clouds. { Polaris is shown but not included in this analysis due to the small number of pixels above the extinction boundary as discussed previously (\S\ref{sec:npdf}).} We will examine the mass--size relation for higher boundaries in the next section.

We derive the slope $a$ and coefficient $C$ for the mass-{size} relation, $M = C R^a$, using a linear regression fit in log space, $\log{M} = a \log{R} + \log{C}$, with mass $M$ given in \msun, radius $R$ in parsecs, and coefficient $C$ in units of ($\msun\ {\rm pc}^{-a}$). In this expression, if $a\to2$, then $C\to\Sigma \pi$. We find $M = 94\, R^{2.07}$ using extinction and $M = 61\, R^{2.22}$ for the CO data.

For the extinction measurements, it is instructive to first compare our results to that of  \citetalias{2010A&A...519L...7L} since both studies are based on measurements of the dust to determine physical cloud properties. \citetalias{2010A&A...519L...7L} use near-infrared dust extinction measurements of background stars to derive the masses and sizes of the GMCs, while here we use extinction calibrated measurements of dust emission, as described earlier. Generally, dust emission studies can probe higher extinction regions than dust extinction measurements; although, such high extinction regions account for a small fraction of the area of a GMC. The angular resolutions of the two studies differ considerably. For the \citetalias{2010A&A...519L...7L} study, the angular resolution ranges between 1\farcm3--3\arcmin\  while for this work it is significantly worse, $\sim$ 8\farcm6. The sizes of the samples are nearly identical (11 vs 12 GMCs) with the targeted clouds having significant overlap. Specifically, the Orion A, Orion B, Taurus, Perseus, Ophiuchus, California, and RCrA clouds are included in both samples and account for slightly more than half of each one. %

Our fit to the dust-derived mass--size relation shown in Figure \ref{fig:massradius} is in excellent agreement with that derived from the \citetalias{2010A&A...519L...7L} measurements:  $M=129\, R^{1.99}$ for $A_0 = 0.1\ \magn$. In particular, the slopes of the relations are essentially identical within the uncertainties and very close to a value of 2, indicating that local GMCs are indeed described by a constant average column density. Indeed, averaging the individual GMC surface densities in each sample gives $\avg\Sigma = 38 \pm 8\ {\rm and}\  41 \pm 5$ \msunpc for the present work and that of \citetalias{2010A&A...519L...7L}, respectively. However, the dispersion in the surface densities of the \citetalias{2010A&A...519L...7L} sample is about a factor of  2 lower ($\sim$ 10$\%$ vs. 20$\%$) than that characterizing the sample in this paper. The origin of this difference is not clear but may be related to the different resolutions of the two studies or to the different clouds contained in the two samples. If we take only those (seven) sources that are the same in both studies, we find $\avg\Sigma = 39.3\pm6.0\ \msunpc$ for the present study and $\avg{\Sigma} = 39.6\pm3.9\ \msunpc$ for \citetalias{2010A&A...519L...7L}, approximately halving the difference between the scatter in the two studies. Additionally, the error we measure for $\avg\Sigma$ ($\sim20\%$) for the full sample is considerably smaller than the error predicted for a sample of clouds with varying $n$ ($\sim30\%$). We will discuss this discrepancy in the next section.

The CO derived mass--size relation shown in Figure \ref{fig:massradius} does differ somewhat from the dust derived relation. In particular, the  slope ($a$) of the CO derived relation is steeper (2.22) than that of the dust-derived relation (2.07), and this difference appears to be marginally significant. The coefficient ($C$) is also found to be different for the two fits.   
Despite the differences in  $a$ and $C$, we do directly measure a similar average surface density \avgsig$\approx$37--38\msunpc for the clouds plotted in the two relations. Overall, the CO and dust-derived mass--size relations are in reasonably good agreement, indicating that the two methodologies do return similar results when the mass--size relations are derived in a systematic fashion using fixed boundaries for the clouds.

We note that the mean surface densities we measure are similar to the average surface density ($\langle\Sigma\rangle=35\msunpc$) measured for GMCs outside the molecular ring in the Milky Way disk  by \citetalias{2020ApJ...898....3L} using CO. Their CO measurements, much like ours, were rescaled to a fiducial {\it X}-factor calibration measured through comparison with surface densities from the \citet{2010ApJ...724..687L} sample of local clouds, which were also defined with $\A0=0.1$ mag boundaries.

\subsection{Linking the Mass--Size Relation with the PDF}\label{sec:mass_size2}
As discussed in Section \ref{sec:mass_size0} and shown by Equation \eqref{eqn:avgext}, the average measured surface density of a GMC with a power-law PDF is directly proportional to the boundary column density used to define the cloud. The mass and area of cloud can be derived directly from integrals of the PDF, as discussed in \citetalias{2010A&A...519L...7L}. The surface density is the ratio of the mass and area, leading to a function proportional to Equation \eqref{eqn:avgext}, $\Sigma_{\rm GMC} = \sigcon \avg\ak$.
In Table \ref{tab:massradius}, we show results of linear fits to both the extinction and CO mass--size relations for {our ensemble of clouds, with} varying extinction and CO boundary levels, $A_0$ and $W_0$, respectively. For the extinction mass-{size} relation, comparison with Table \ref{tab:pdf} shows that the lowest boundary is below the lowest closed contour ($\akmin$) in the clouds. A measured slope of $a\approx2$ for the extinction derived mass--size relation is the expected result for a sample of clouds that have a constant or near constant surface density, provided that the same fixed boundary is used to calculate the average surface density for all clouds being compared. 
Changing the boundary has a significant effect on the value of $C$ and \avgsig, as expected, from Equation \eqref{eqn:avgext}. 
\begin{deluxetable}{rCcCcc}
\tablecaption{Mass--Size Relation}
\label{tab:massradius}
\tablehead{\colhead{Limit ($\Sigma_0$)} & \colhead{Slope ($a$)} & \colhead{$C$} & \colhead{$\langle\Sigma\rangle$} & \colhead{$\langle\Sigma\rangle/\Sigma_0$} & \colhead{scatter}}
\startdata
\cutinhead{\A0 [mag]}
0.03 & 2.36 \pm 0.13 &   27  &   28 \pm  8  &  5.1  &   15\% \\
 0.1 & 2.07 \pm 0.09 &   94  &   38 \pm  8  &  2.1  &   16\% \\
 0.2 & 2.02 \pm 0.07 &  182  &   62 \pm 12  &  1.7  &   15\% \\
 0.3 & 2.03 \pm 0.07 &  254  &   87 \pm 18  &  1.6  &   15\% \\
 0.5 & 1.98 \pm 0.08 &  434  &  137 \pm 19  &  1.5  &   11\% \\
\cutinhead{\W0 [\Kkms]}
1.5 & 2.22 \pm 0.12 &   61  &   37 \pm 10  &  5.8  &   16\% \\
 4.5 & 2.18 \pm 0.11 &  103  &   53 \pm 14  &  2.7  &   16\% \\
   9 & 2.12 \pm 0.09 &  171  &   72 \pm 17  &  1.9  &   16\% \\
13.5 & 2.11 \pm 0.07 &  230  &   92 \pm 19  &  1.6  &   14\% \\
  22 & 2.10 \pm 0.04 &  360  &  130 \pm 22  &  1.4  &   10\%
\enddata
\tablecomments{Results from linear regression fit to the mass--size relation. $\Sigma_0$ is the boundary, $C$ is a scaling constant ($C=\Sigma \pi$ for $p=2$), and $\langle\Sigma\rangle/\Sigma_0$ is the ratio of the average surface density to the boundary level. $\langle\Sigma\rangle/\Sigma_0 = 2$, corresponds to an extinction PDF slope $n=3$.}
\end{deluxetable}

The CO mass--size relation displays remarkably similar behavior to the extinction relation in the constancy of its slopes with increasing boundary or threshold levels and the clear boundary dependence of the normalizing constant $C$ and $\avg\Sigma$, as expected from Equation \eqref{eqn:avgext}. Moreover, the dispersion in \avgsig\  is relatively small ranging from $\sim$ 17-27 $\%$. This is, at first glance, surprising given how poorly the CO PDFs correspond to single power-law functions. However,  \citet{2012MNRAS.427.2562B} have shown that such behavior is not solely confined to single power-law PDFs such as those characterized by Equation \eqref{eqn:avgext}, but it also characterizes any steeply falling PDF, regardless of its exact functional form. This is a result of the simple consideration that for a PDF that is steeply falling with surface density in log-log space, most of the cloud material must have surface densities near the boundary  value and the average surface density will thus be not too far removed from the value of the threshold boundary. 

The CO mass--size relation does differ in some respects from the extinction relation.  Notably, the slope derived using the CO data is consistently higher than that ($\sim$ 2.0) of the extinction relation. 
The slope of $a\sim2.2$ is similar to what was measured for the \citet{2017ApJ...834...57M} Galactic CO cloud catalog by \citetalias{2020ApJ...898....3L}. \citet{2019MNRAS.490.2648B} suggest that superposition of unrelated emission from clouds overlapping along the line-of-sight could cause $a$ to increase to $2.2-2.3$; however, our clouds are all nearby and selected to be isolated in velocity space. If superposition were a significant problem resulting in increased $a$, then we would expect to see this effect in the dust measurements of our sample GMCs, where we know there is a small amount of contaminating material that we have not accounted for. In comparing the CO and extinction mass--size relations, it is important to note that due to variations in \xco within a cloud, a fixed \wco boundary does not correspond to any single extinction boundary. 
In general, converting a boundary using \avg\aco and \sigcon will result in regions with $\lesssim70\%$ overlap. The difference in the area covered likely contributes to the different slopes as well. The physical origin of the slightly larger slope for the CO mass-{size} relation is unclear. It may be related to the fact that the CO PDFs are not simple power laws as the extinction PDFs are, but more detailed tests need to be conducted before any firm conclusions can be drawn.

\begin{figure*}
    \centering
    \plottwo{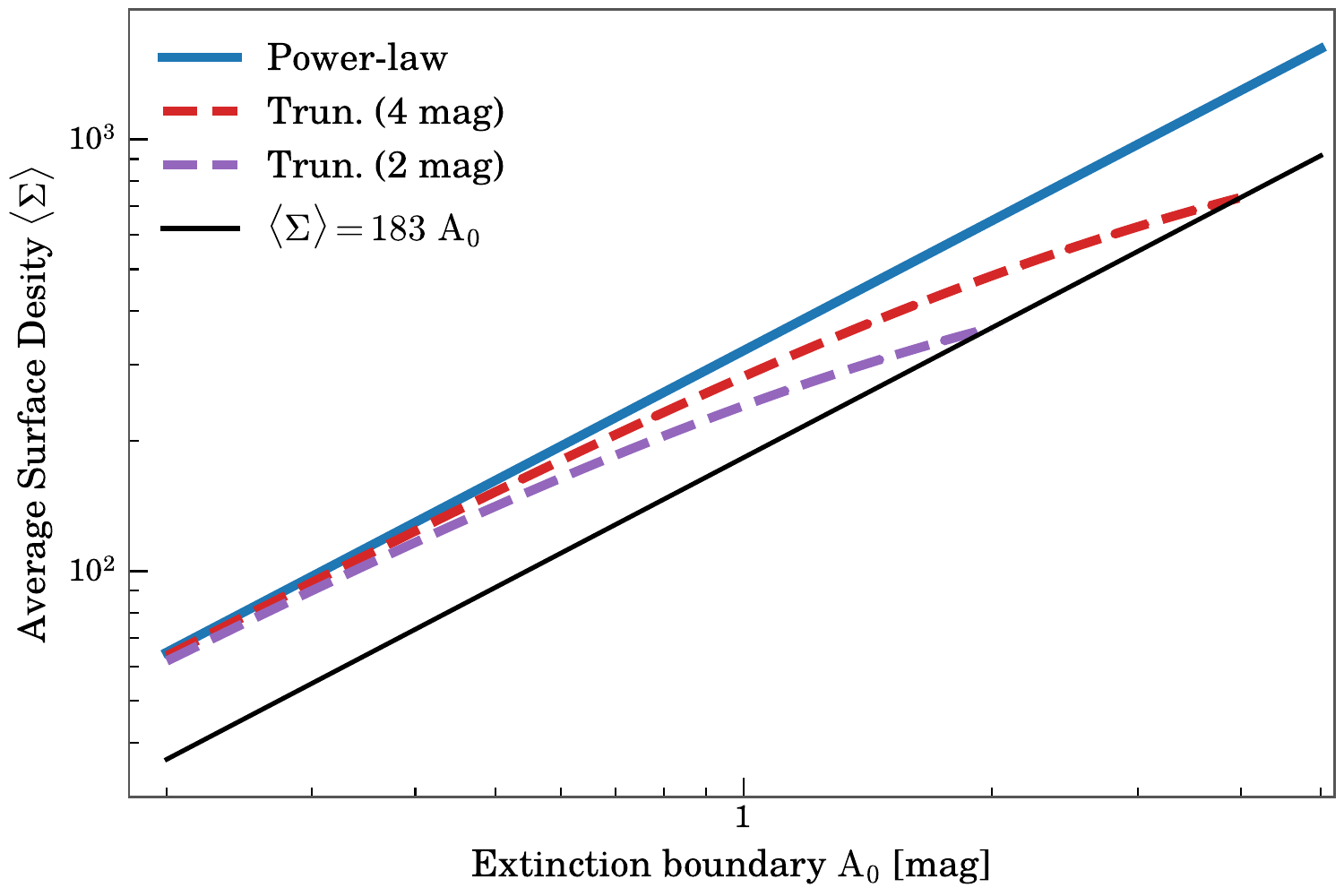}{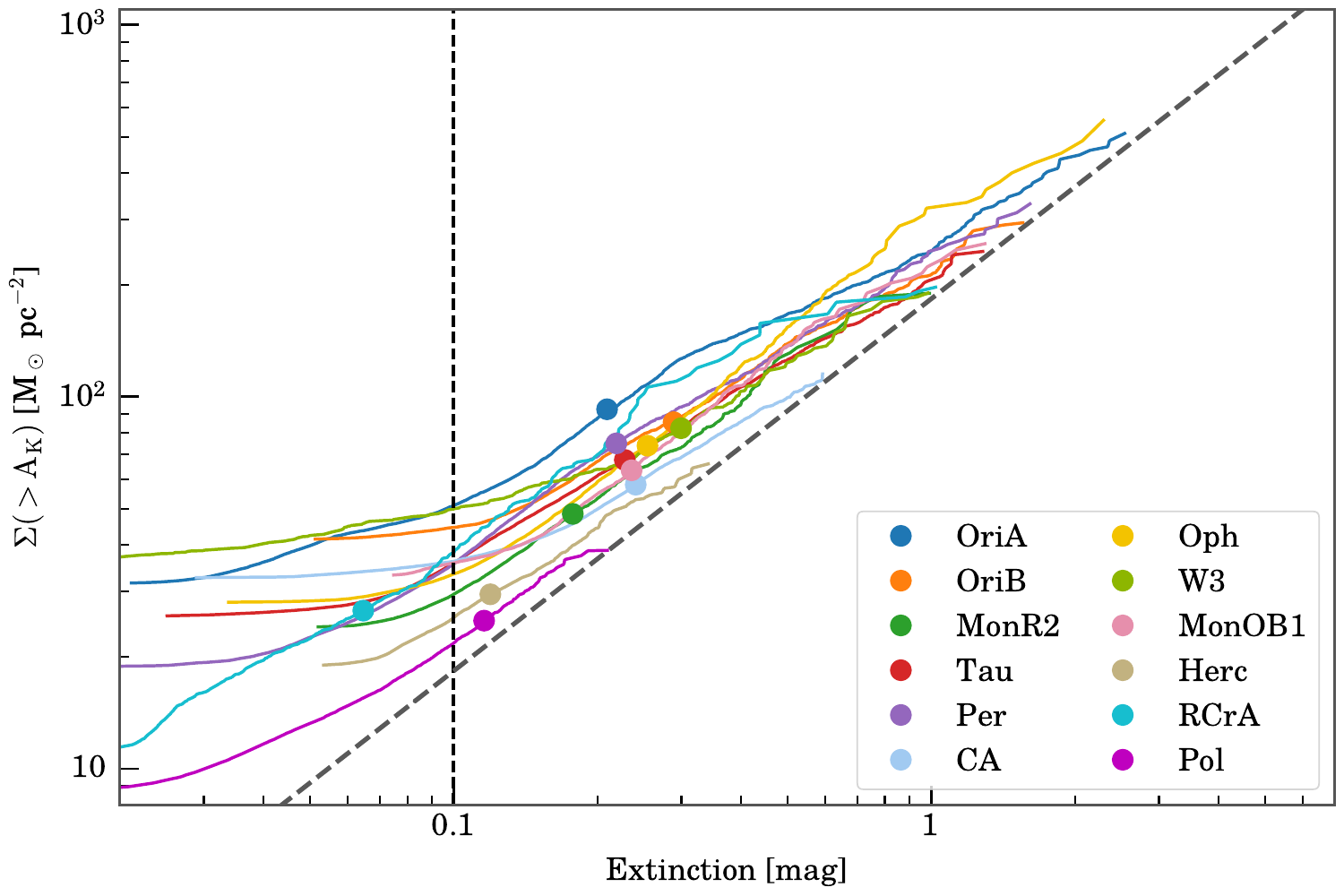}
    \caption{Left panel: theoretical curves for average surface density as a function of threshold extinction.  This example is for a cloud with power-law slope $n=3.3$, $\akmin=0.2$, and $\akmax= $ 2 or 4 mag. 
    Right panel: surface density vs extinction threshold for all the molecular clouds using the $8\farcm6$ \planck data. The points are located at the lowest closed contour value (\akmin) for each cloud.
    }
    \label{fig:cartoon}
    \label{fig:sigmaprofile}
\end{figure*}

From Equation \eqref{eqn:avgext}, we expect the ratio $\avg\Sigma/\Sigma_0$ to be constant. However, Table \ref{tab:massradius} shows $\avg\Sigma/\Sigma_0$ decreasing as the cloud boundary rises, for both extinction and CO boundaries. Here, we show that this result is due to the fact that in the observations, the PDFs are often truncated at the high extinction end of the PDF due to limitations of angular resolution.
In Appendix \ref{app:derivatin}, we show that for a power law truncated at the peak extinction \akmax, the surface density no longer has a linear relation with the boundary at all extinctions. The surface density of a truncated distribution is, for
$\A0 > \akmin$,
\begin{linenomath}
\begin{equation}
    \begin{split}\label{eqn:truncsigma}
    \Sigma_T(>\!\A0) &= \sigcon \frac{n-1}{n-2} f(\A0)  \A0, \\
                    & f(\A0) = \left( \frac{\A0^{1-n} - \A0^{-1}\akmax^{2-n}}{\A0^{1-n} - \akmax^{1-n}} \right)
    \end{split}
\end{equation}
\end{linenomath}
This more complex equation shows that if the range of extinctions for which a power law is valid is bounded both above and below, then the surface density will not be linearly related to the boundary extinction over all values of extinction. This is always the case for our real data. The term $f(\A0)$ is $<1$ for $\A0\in(0,\akmax)$ and $f(\A0)\to\frac{n-2}{n-1}$ as $\A0\to\akmax$. This means that $\Sigma/\Sigma_0$ for a cloud will not be a constant across the full range of extinction, but will approach $1$ as \A0 increases. Also note that this is independent of \akmin. In Figure \ref{fig:cartoon}, we show the average surface density for a simple power law (blue) and two example truncated power laws (red and purple) for a cloud with $n = 3.3$ and $\akmax = $ 2 (purple) or 4 mag (red). The effect of truncation is evident. Within a single cloud, a high boundary near the peak map value will underestimate the true surface density.   For a collection of clouds, where \akmax and $n$ varies, this means $\langle\Sigma\rangle/\Sigma_0$ decreases as \A0 increases. This is what is seen in Table \ref{tab:massradius} for $\langle\Sigma\rangle/\Sigma_0$.

In Figure \ref{fig:sigmaprofile}, we examine the surface density as a function of the boundary. When using the parameters from Table \ref{tab:pdf}, our molecular clouds follow Equation \eqref{eqn:truncsigma} to within 6\%. In the previous section (\S\ref{sec:constsurf}), we noted that the fractional error in $\avg\Sigma$ was smaller than expected from Equation \eqref{eqn:avgext}. If we use Equation \eqref{eqn:truncsigma} to estimate the expected scatter including the different values of $n$ and \avg\akmax for \akmax, then we find a fractional error of 18\%, in line with what we see in Table \ref{tab:massradius}. In Appendix \ref{app:derivatin}, we show this graphically. Above the closed contour, the scatter in $\avg\Sigma$ for a single cloud boundary is dominated by scatter in $n$. By accounting for the upper truncation of the PDF, we can account for the deviation of \avgsig/$\Sigma_0$ from being strictly constant. Higher resolution dust measurements push \akmax to higher values. In the cold California cloud, \citet{2017A&A...606A.100L} showed a power-law extending to the highest column densities measured. This shows, that for a single cloud boundary, the molecular clouds in our sample do truly have a  constant surface density to a high degree ($\sim10\%$) of precision. 

\subsection{Universality of GMC Surface Densities}

The average column or surface density of a GMC population can be an important metric for characterizing cloud properties in comparative studies of GMCs across and between galaxies.
The dust observations analyzed in this paper robustly confirm results of earlier work that indicated a relatively precise constancy of the average surface densities of local Milky Way clouds. The explanation of this constancy has been directly linked to the steeply falling nature of the cloud PDFs \citep{2012MNRAS.423.2579B, 2012MNRAS.427.2562B}.  Indeed, as shown in Section \ref{sec:npdf}, power-law PDFs explicitly require the average surface densities to be constant, as is clearly demonstrated from observations of the local cloud sample. Moreover, when measured from a fixed surface density boundary, the average cloud surface density ($\Sigma_{\rm GMC}$ ) depends weakly on the value of the index of the power law for n $>$ 2. This implies that the specific value of $\Sigma_{\rm GMC}$ should be universal for similarly constructed GMCs. Moreover, this value is expected to be ~ 35-45\msunpc if measured relative to a fixed boundary that is close to \ak $=$ 0.1 mag or \wco $\approx$ 1.5 K km/s. Interestingly, this seems to be the case for clouds in the disks of numerous nearby galaxies \citep{2008AJ....136.2846B,2018ApJ...857...19F,2020ApJ...901L...8S}, where the average surface densities derived from CO observations are observed to rarely exceed $\sim$50-60\msunpc outside their nuclear regions.  Moreover, it has been suggested that the surface densities that are measured to be lower than $\sim$ 50-60 M$_\odot$ are likely beam diluted \citep{2008AJ....136.2846B,2015IAUS..309...31L} and thus have intrinsically higher values, which is consistent with the notion of a universal $\langle\Sigma_{\rm GMC}\rangle$ in these galaxies. The difference between the slightly larger value  of $\langle\Sigma_{\rm GMC}\rangle$ (50-60\msunpc) obtained for galaxies compared to that (35-45\msunpc) found for local clouds is likely due to differences in the effective boundaries used for the different measurements, and probably not significant. The observed similarity for cloud PDFs is not without theoretical support because of the turbulent nature of molecular clouds. Simulations universally show that molecular clouds formed from a turbulent ISM are characterized initially by log-normal PDFs that evolve to power laws once gravity becomes important  \citep[e.g.;][]{1997ApJ...474..730P,2011MNRAS.416.1436B,2013ApJ...763...51F,2017ApJ...834L...1B}.

However, there are notable exceptions to a universal $\langle\Sigma_{\rm GMC}\rangle$, such as the galaxy M51 where the median $\Sigma_{\rm GMC}$ was found to be $\approx$ 180\msunpc \citep{2014ApJ...784....3C}.  It is possible some of this difference is due to the  use of effectively higher boundaries in the measurements of M51 surface densities. For example, in M51 clouds must be detected above a significant (4 K) diffuse molecular component, which could increase the surface density of the cloud boundary used in the measurements. However, it may also be possible that the environment in M51 is sufficiently different from the Milky Way and most spiral galaxies that the nature and internal structure of the clouds has been modified. Equation \eqref{eqn:avgext} and Figure \ref{fig:surfind} show that $\langle\Sigma_{\rm GMC}\rangle$ is very sensitive to $n$ if $n$ has a value close to 2. So, is it possible that the environment in this galaxy has produced molecular clouds with flatter PDFs and higher fractions of high density material? The internal pressure of a cloud is proportional to the square of its surface density ($P_{\rm GMC} \propto G \Sigma^2_{\rm GMC}$, \citealt{1992ApJ...395..140B}). If GMCs are in pressure equilibrium with their surroundings, then their internal pressure would be directly related to the pressure of the surrounding ISM. Indeed, \citet{2018ApJ...857...19F} recently showed that the ratios of  midplane pressures between M51, the Milky Way and NGC 300 are roughly equal to the ratios of the squares of the average cloud surface densities of these galaxies. These authors postulated that the different measured surface densities in the cloud populations of these galaxies were due to the higher midplane ISM pressure of M51 compared to the Milky Way and NGC 300, two galaxies with similar midplane  pressures. Cloud pressure has been found to be correlated with average cloud surface density in a large sample of nearby galaxies, \citep[e.g.,][]{2020ApJ...901L...8S} so a link between ISM pressure and GMC surface density seems to be indicated by recent observations. The potential role of pressure in influencing molecular cloud properties has long been suggested \citep{1989ApJ...338..178E,2004ApJ...612L..29B}. Evidence for departures from a universal $\langle\Sigma_{\rm GMC}\rangle$ may be also present in the Milky Way disk. As mentioned earlier, \citetalias{2020ApJ...898....3L} found a radial dependence for $\langle\Sigma_{\rm GMC}\rangle$ in the Galaxy. Within the molecular ring, the average surface densities were found to be a factor of $\sim$ 2 higher than in the inner and outer regions of the Milky Way disk. However, they argued that the cloud surface density measurements in the molecular ring were biased, due to the effects of cloud overlap and blending as seen from earth; consequently, the degree to which they depart from a disk wide constant $\langle\Sigma_{\rm GMC}\rangle$ is very uncertain. However, the radial variation in the midplane pressure of the Milky Way might still produce a smaller increase in $\langle\Sigma_{\rm GMC}\rangle$ of about  50-70\% within the molecular ring \citepalias{2020ApJ...898....3L}, resulting in a departure from a universal value in that region.   Finally, we note that average cloud  surface densities can range from 100-1000\msunpc in the nuclear regions of some galaxies, such as starburst galaxies and barred spirals \citep{2015ApJ...803...16U,2020ApJ...901L...8S}. The highest of these surface densities would be difficult to explain in the framework outlined above. The clouds in these nuclear regions are clearly of a significantly different nature than clouds considered here. The nuclear GMCs are typically much more massive, have much higher velocity dispersions and are located in regions with strong tidal forces. These GMCs are likely different in nature and not governed by the same scaling laws that describe disk clouds.

\section{Summary}\label{sec:summary}
In this paper, we report the results of the first uniform and systematic comparison of the molecular gas and dust in  local GMCs. We analyzed data from existing infrared surveys of dust extinction and emission to determine and define the basic physical properties of the clouds in a self-consistent manner. We also analyzed data from a $J=$1-0 CO survey of the Galactic plane in a similar self-consistent manner as the dust analysis and compared the results. From this comparison we derived $\alpha$, the CO mass conversion factor, and \xco, the CO-to-H$_2$ conversion factor for the cloud sample. We also evaluated the efficacy of CO observations for measuring the basic properties of the clouds.  Below we summarize the primary findings of the paper.

\begin{itemize}
    \item In local clouds, the optical depth ($\tau_{353}$) to extinction (\ak) conversion factor $\gamma = \ak/\tau_{353}$ spans a factor of $\sim$2 in range (2300-4500), with $\avg\gamma = 3350 \pm 620$. The large variation implies that using a single dust opacity ($\kappa$), as is common, will lead to systematic errors of order 20\% in the estimation of molecular cloud mass. 
    
    \item We show that $\gamma$ is related to the dust opacity. Assuming a dust to gas mass ratio, we find that an average $\gamma$ $\sim$3300 corresponds to $\kappa_{\rm d, 353} = 0.8\ {\rm cm^2\ g^{-1}}$. This value lies between what is expected for the diffuse ISM and ISM in protostellar cores. 
    
    \item We calculate the CO luminosity to total gas mass conversion factor, \aco, and find it to range from about $2.7-8\ \alphaunit)$ in our sample. 
    For the sample average we find $\langle\aco\rangle$ =
    $ 4.31 \pm 0.67\ \alphaunit$ (excluding Ophiuchus and Polaris)
    corresponding to \xco $=$ 1.99\E{20}\xfunit, essentially the values suggested by \citet{2013ARA&A..51..207B}. 
    
    \item We measured \xco, the conversion factor between CO integrated intensity (\wco) and H$_2$ column density, on sub-cloud (parsec level) scales. On these scales we find a significant amount of  variability  in \xco that is characterized by a broad log-normal-like frequency distribution, indicating that CO is a poor tracer of H$_2$ column density on these scales.

    \item We find that the internal structures of individual GMCs are characterized in the dust observations by simple power-law PDFs to relatively high ($\sim$ 10\%) precision. These power-laws span a range in extinction that extends from our completeness limits (\ak $\sim$ 0.1--0.2 magnitudes) to the highest measurable values in the individual clouds, confirming an earlier study. The average spectral index for our local cloud sample is $\langle n\rangle = 3.66\pm0.74$. RCrA, the most isolated cloud in the sample, shows a power-law profile down to 0.03 mag, very near the \HI-\htwo cloud interface. 
    
    \item We find the PDFs of CO integrated intensities not to be characterized by simple, well-defined power-law functions in the individual GMCs.  This once more reflects the fact that CO is a poor tracer of column density on sub-cloud scales. However, similar to the dust, the CO PDFs are still found to generally decrease with integrated CO intensity.

    \item We measure the mass--size scaling relation for GMCs in our sample using the dust observations and find $M\sim R^{2.07}$ indicating that the clouds are characterized by a constant surface density to relatively high precision above some boundary, $\Sigma_0$, and we find  $\avg\Sigma = 38 \pm 8\ \msunpc$ for $\Sigma_0$ = 0.1 magnitudes. The tight constancy of cloud surface densities is expected for clouds with a power-law PDFs characterized by a similar index ($n$).  We further show that the measured dispersion in $\avg\Sigma$ may likely be due to the dispersion in the indices of the cloud PDFs coupled with the effects of the observed truncation of the PDFs at high extinction. 
    
    \item We find that, except at the highest extinctions measured in a cloud, the derived value of $\langle \Sigma \rangle$ depends linearly on $\Sigma_0$, as predicted for clouds characterized by power-law PDFs. At the highest extinctions we observe a departure from strict linearity and show that this is the result of the truncation of the PDFs resulting from the limited angular resolution of the observations. These results provide strong additional support for a constancy of the average surface densities of local GMCs. Moreover, these considerations clearly demonstrate that care must be taken to adopt common fixed boundary levels when comparing surface densities of different cloud populations.

\end{itemize}

One of the goals of this paper was to assess the comparative abilities of dust and CO observations to measure the basic physical properties of molecular clouds. Observations of the dust are very effective tracers of H$_2$ and consistently provide robust measures of GMC properties within and between clouds.  Despite the inability of CO to accurately trace H$_2$ on sub-cloud scales, the GMC mass--size relation derived from the CO observations is only slightly steeper (i.e., $M\sim R^{2.22}$) than the dust-derived relation. However, the small dispersion in the 
sample averaged surface density ($\avg\Sigma$), measured relative to a fixed boundary (e.g., $\langle \Sigma \rangle = 37 \pm 10$ M$_\odot$ pc$^{-2}$ for $W_0$ = 1.5 K km s$^{-1}$), is consistent with a constant average surface density for our GMC sample. We conclude that, although not as robust as observations of dust, CO observations calibrated by dust measurements (i.e., \aco) can be effectively used to measure and compare bulk cloud properties, such as the average cloud surface density, when such measurements are referenced to a fixed common cloud boundary.

\acknowledgments
We thank the referee Javier Ballesteros-Paredes for thoughtful comments that improved the paper.

\vspace{5mm}
\facilities{CfA 1.2m Millimeter Telescope, Planck}

\software{astropy \citep{2013A&A...558A..33A, 2018AJ....156..123A},
          healpy \citep{2005ApJ...622..759G,2019JOSS....4.1298Z},
          Matplotlib \citep{Hunter2007,thomas_a_caswell_2021_5706396},
          SciPy \citep{2020NatMe..17..261V},
          Montage/MontagePy \citep{2010arXiv1005.4454J,2010ascl.soft10036J},
          powerlaw \citep{2014PLoSO...985777A}
          }

\bibliography{library,other_bib}{}
\bibliographystyle{aasjournal}

\appendix
\section{The Mass -- Size relation for a truncated power-law}\label{app:derivatin}

In this appendix, we will derive and analyze the truncated form of the power-law distribution seen in real data. \citet{2012MNRAS.427.2562B} showed that a power-law PDF naturally gives rise to a $M\sim R^a$ mass-{size} relation where $a=2$. Briefly, for any probability distribution describing a molecular cloud the mass and size of the cloud can be derived directly from the PDF $p(A_{\rm K})$ using the equations,
\begin{linenomath}
\begin{align}
    S(>\!\A0) &=  S_{\rm tot} \int_{\A0}^\infty p(\ak^\prime)\ \dd \ak^\prime \label{eqn:sizeint}\\
    M(>\!\A0) &= \sigcon S_{\rm tot}  \int_{\A0}^\infty \ak^\prime \, p(\ak^\prime)\ \dd \ak^\prime \label{eqn:dmassint}, 
\end{align}
\end{linenomath}
where $S_{\rm tot}$ is the total area of the cloud, and $\sigcon=\mu m_p \beta_K$ is the conversion from extinction to \msunpc. The surface density is simply $\Sigma(>\!\A0)\equiv \frac{M(>\!\A0)}{S(>\!\A0)}$.  Because we use a different notation, we briefly describe the results for a power-law PDF (eqn. \ref{eqn:plpdf}).
The area, mass, and surface density are
\begin{linenomath}
\begin{align}
    \textstyle S(>\!\A0) &=\textstyle S_{\rm tot} \left(\frac{\A0}{\akmin}\right)^{1-n} \label{eqn:darea}, \\ 
    \textstyle M(>\!\A0) &=\textstyle \sigcon S_{\rm tot} \frac{n-1}{n-2}\, \left(\frac{\A0^{2-n}}{\akmin^{1-n}} \right) \label{eqn:dmass}, \\ 
    \textstyle \Sigma(>\!\A0) &=\textstyle \sigcon \frac{n-1}{n-2}\A0, \label{eqn:dsurf}
\end{align}
\end{linenomath}
Because $\Sigma(>\!\A0)$ is constant, this implies a $M\sim R^2$ relation between clouds with similar $n$. As \citeauthor{2012MNRAS.427.2562B} also pointed out, this does not imply that the $M\sim R^2$ holds within clouds. In fact, within clouds the mass-{size} relationship depends on the PDF slope. If $M\sim R^b$, then within power-law PDF clouds, 
$ b = 2\frac{\dd\log{M}}{\dd\log{S}} = 2 \frac{n-2}{n-1} $,
which is strictly $<2$. 

What neither the power-law description from \citeauthor{2012MNRAS.427.2562B} nor the log-normal formulation from \citet{2010A&A...519L...7L} really capture are the change in $\avg\Sigma/\A0$ with \A0 and the turn-over of the intra-cloud mass-{size} relation for large \A0. As shown in Table \ref{tab:massradius}, the ratio $\avg\Sigma /\A0$ decreases with \A0. \citet{2010A&A...519L...7L} suggest such effects could be due to limited dynamic range. We show that a truncated power-law PDF is a better description of the data than a standard power-law and reproduces the effects seen here and in \citeauthor{2010A&A...519L...7L}.

\begin{figure}[th]
    \centering
    \plotone{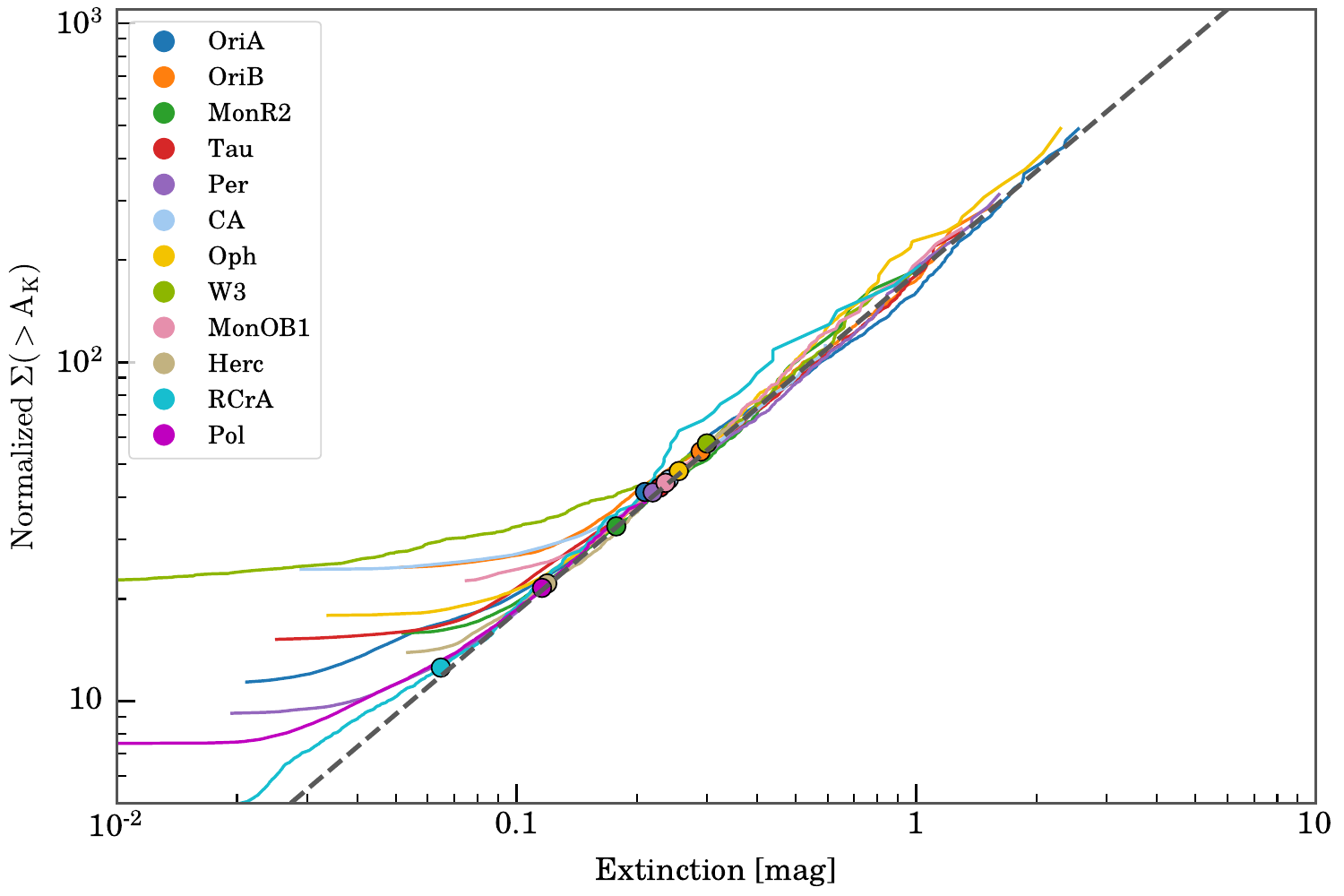}
    \caption{Normalized surface density profiles $\Sigma\  (f(\A0)\frac{n-1}{n-2})^{-1}$}
    \label{fig:mrprofile}
\end{figure}

The equation for a power-law truncated at some peak extinction $A_{\rm K, max}$, $p_{\rm T}(\ak)$, is essentially that same as the normal power-law, except it is normalized so the integral from \akmin$\to$\akmax is equal to one,
\begin{linenomath}
\begin{equation}
    p_{\rm T}(\ak) = \frac{\akmax^{n-1}}{\akmax^{n-1}-\akmin^{n-1}} p(\ak)\  \forall\ \ak \in [\akmin,\akmax].
\end{equation}
\end{linenomath}
Here $p(\ak)$ is the standard power-law PDF defined in equation \ref{eqn:plpdf}. Table \ref{tab:pdf} lists $\akmax$ for each cloud. The equations for the mass and size functions have the same integral form as previously given in equations (\ref{eqn:sizeint}) and (\ref{eqn:dmassint}), but now are evaluated over the range  \A0 to \akmax. For $p_{\rm T}$, the mass and size functions are
\begin{linenomath}
\begin{align}
    S_T(>\!\A0) &= S_{\rm tot} \frac{\akmax^{n-1} - \A0^{n-1}}{\akmax^{n-1}-\akmin^{n-1}} \left(\frac{\A0}{\akmin}\right)^{1-n} \\ 
    M_T(>\!\A0) &= \sigcon S_{\rm tot} \frac{n-1}{n-2}\, \frac{\akmax^n\, \akmin^n}{\akmax^n\,\akmin - \akmin^n\,\akmax}\left(\A0^{2-n} - \akmax^{2-n}\right),
\end{align}
\end{linenomath}
where $\A0\in[\akmin,\akmax)$ and $n>2$. The surface density $\Sigma_T = M_T / S_T$, after some rearranging is,
\begin{linenomath}
\begin{align}
    \Sigma_T(>\!\A0) &= \sigcon \frac{n-1}{n-2} f(\A0)  \A0, \label{eqn:appsigma}\\
                    & f(\A0) = \left( \frac{\A0^{1-n} - \A0^{-1}\akmax^{2-n}}{\A0^{1-n} - \akmax^{1-n}} \right) \nonumber
\end{align}
\end{linenomath}
This more complex equation shows that due to limited dynamic range the surface density in a cloud will not be linearly related to the boundary chosen to measure it. The term $f(\A0)<1$ for $\A0\in(0,\akmax)$ and $f(\A0)\to\frac{n-2}{n-1}$ as $\A0\to\akmax$. The means that $\Sigma/\Sigma_0$ for a cloud will not be a constant, but will approach $1$ as \A0 increases. For a collection of clouds, where \akmax and $n$ varies, this means $\langle\Sigma\rangle/\Sigma_0$ $\downarrow$ as \A0$\uparrow$. As discussed in \S\ref{sec:mass_size2}, and shown in Figure \ref{fig:cartoon}, the shape of the $\Sigma$ -- \A0 profile closely follows the prediction from equation \ref{eqn:appsigma}. To show how well, in Figure \ref{fig:mrprofile}, we normalize the $\Sigma$ -- \A0 profiles for all the clouds, plotting $\frac{\Sigma}{ f(\A0)\frac{n-1}{n-2}}$ vs \A0. The relative scatter for all points $>0.2$ mag is only 6\%. Before correction, the scatter in $\langle \Sigma \rangle$ was $\sim 20\%$, indicating a large fraction of the scatter comes from real differences between clouds once variable boundaries are dealt with (see \S\ref{sec:mass_size0} and \S\ref{sec:mass_size1}).

\newpage
\section{Log-normal X-factor Distribution}\label{app:xco}
In Section \ref{sec:cointro}, we used our data to measure the mass conversion factor  (\aco or equivalently \xco) for CO on the spatial scales of entire GMCs. However, these global conversion factors do not provide accurate measures of gaseous masses on sub-cloud scales because on those scales the X-factor has been shown to have significant variations within molecular clouds (e.g., Kong et al. 2015). In this appendix, we compare the extinctions measured in each pixel of a map with the value of W(CO) for that pixel to directly compute an {\it on the spot} X-factor for each pixel. With such measurements, we can quantify and investigate  variations of the X-factor on sub-cloud scales across individual GMCs.

\begin{figure}
    \centering
    \plotone{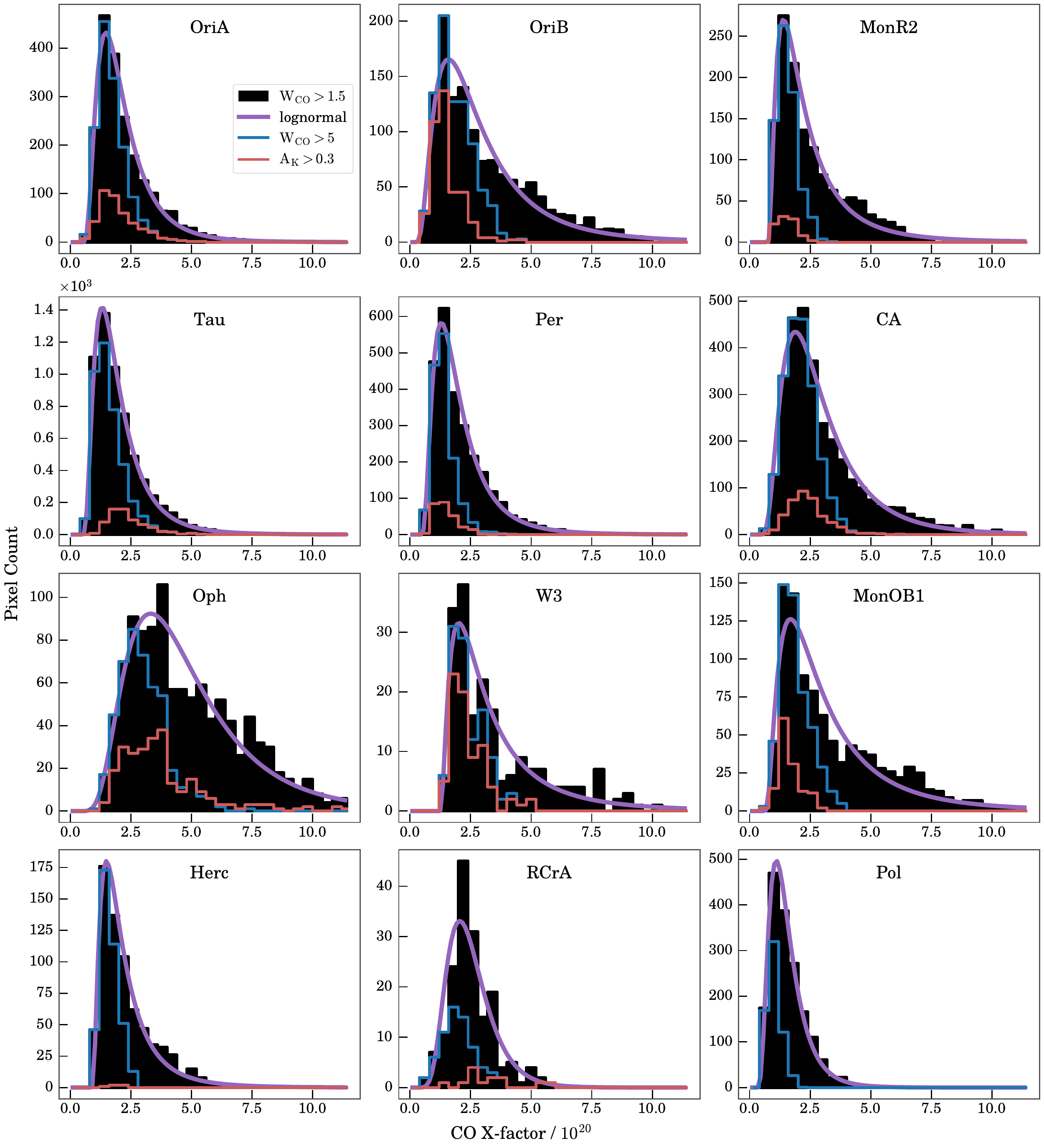}
    \caption{Histogram of \xco. The histogram for pixels where \twelveco is detected (\wco$>1.55\ \Kkms$) is in dark gray. The purple line shows a ML log-normal fit to \xco in those pixels. We show the histogram the results after filtering pixels where $\wco>5\ \Kkms$ and $\ak>0.3\ \magn$ in blue and red, respectively.}
    \label{fig:xcohistperfit}
\end{figure}

Figure \ref{fig:xcohistperfit} shows the frequency distributions of the on the spot  X-factors for each cloud (black trace). These histograms appear to have  log-normal-like shapes, so we fitted them with log-normal functions and plot the fits as a purple line on the individual distributions.  We fitted a shifted log-normal PDF to the histograms of each cloud. The shifted log-normal PDF is given by
\begin{linenomath}
$$p_{\rm L}(X_{\rm CO}) = \frac{1}{(X_{\rm CO}-X_{\rm 0})\sigma_L \sqrt{2\pi}}\exp\!\left(\frac{-\left(\ln(X_{\rm CO}-X_{\rm 0}) - \mu\right)^2 }{2\sigma_L^2}\right),$$
\end{linenomath}
where $X_0$ is a linear translation along the x-axis, and $\mu$ and $\sigma_L$ are the standard parameters of the log-normal distribution, being the mean and standard deviation of $\ln(X_{\rm CO} - X_0)$, respectively. Not including the translation ($X_0$) results in poorer fits with larger $\sigma_L$. Table \ref{tab:lognormfit} shows the parameters of the fit. In RCrA  we force $X_0$ to zero, as the low number of pixels ($N_{pix}]\approx100$) cause the determination of $X_0$ to be noisy and to vary around $X_0=0$. Several clouds---Orion B, Ophiuchus, and Mon OB1---appear to deviate from a log-normal distribution, showing evidence of a multiple peaks.

In Figure \ref{fig:xcohistperfit}, we overplot in blue the histogram of \xco for pixels with $\wco>5\ \Kkms$ and in red the histogram for pixels with $\ak>0.3\ \magn$. In the \wco filtered histogram, the peak stays about the same, but the width narrows by $\sim 40\%$, showing that high \xco values do have a tendency to come from regions with weaker \twelveco lines ($\wco\lesssim 5\ \Kkms$). In the extinction filtered histogram, most of the material comes from lower extinctions. The limit used (0.3 mag) is only $\sim$0.1 mag above the lowest closed contour of most clouds. If we use a lower limit of 0.2 mag (not shown), the \xco distribution above $\xco\sim3$ looks similar to the \wco filtered histogram. The filtered histograms show that the high \xco ($\xco\gtrsim 3$) wing comes from material with low \wco and low extinction (\ak$\lesssim$2-3 mag). This analysis tells us that the {\it X}-factor variations also arise from photodissociation and chemistry on the edges of clouds, even up $\ak\sim0.3$.

\begin{deluxetable}{lccc}
\tablecolumns{4}
\tablewidth{0pt}
\tablecaption{Parameters for Log-normal Fit\label{tab:lognormfit}}
\tablehead{
\colhead{Name} & \colhead{$\sigma_X$} & \colhead{$X_0$} & \colhead{$\mu$}}
\startdata
   Orion A  &  0.62 & 0.47 & 1.46 \\
   Orion B  &  0.70 & 0.18 & 2.31 \\
  Mon R2  &  0.83 & 0.74 & 1.32 \\
    Taurus  &  0.65 & 0.47 & 1.29 \\
    Perseus  &  0.66 & 0.42 & 1.36 \\
     California  &  0.62 & 0.39 & 2.20 \\
    Ophiuchus  &  0.53 & 0.26 & 4.07 \\
     W3  &  0.87 & 1.22 & 1.70 \\
 Mon OB1  &  0.81 & 0.60 & 2.09 \\
   Hercules  &  0.75 & 0.86 & 1.10 \\
   R Corona A  &  0.37 & 0.00 & 2.36 \\
    Polaris  &  0.53 & 0.21 & 1.18
\enddata
\tablecomments{The peak of the distribution is $ \exp{(\mu - \sigma_X)} + X_0$ }
\end{deluxetable}

\newpage
\section{Side by side maps of CO and Dust}\label{app:maps}

\begin{figure}
    \centering
    \plottwo{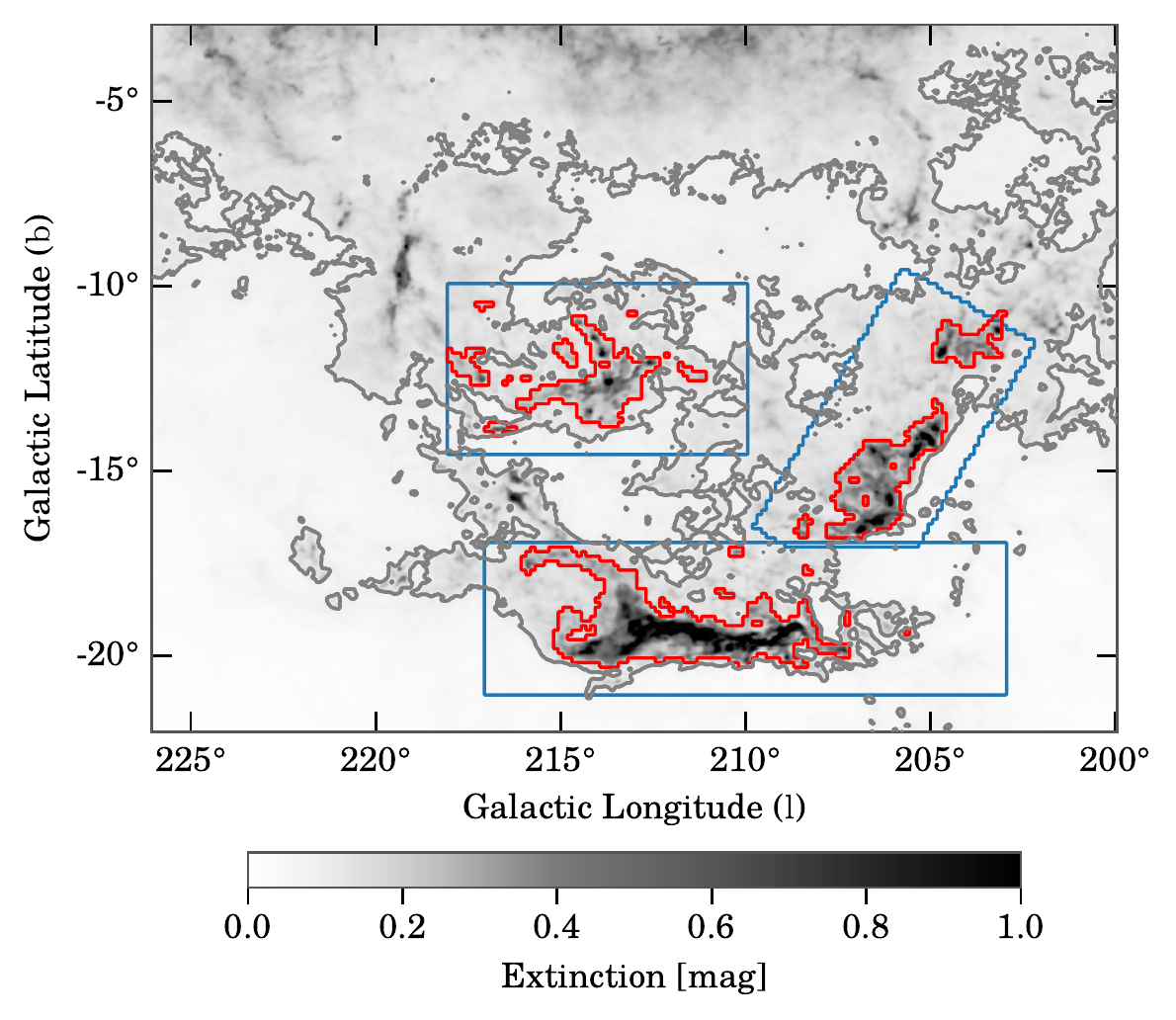}{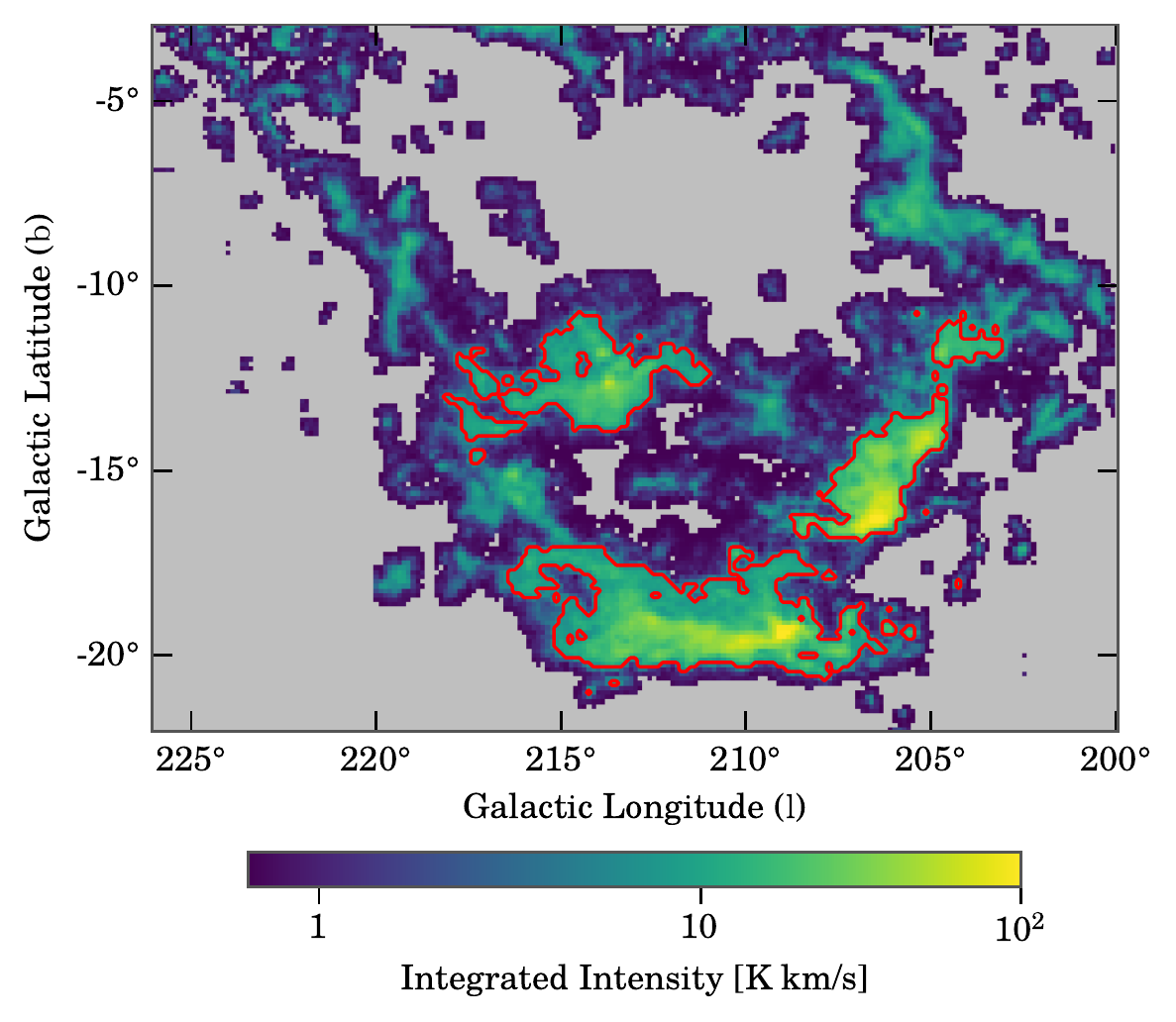}
    \caption{Orion A/B and Mon R2 --- }
    \label{fig:OriA_app}
    \end{figure}

\begin{figure}
    \centering
    \plottwo{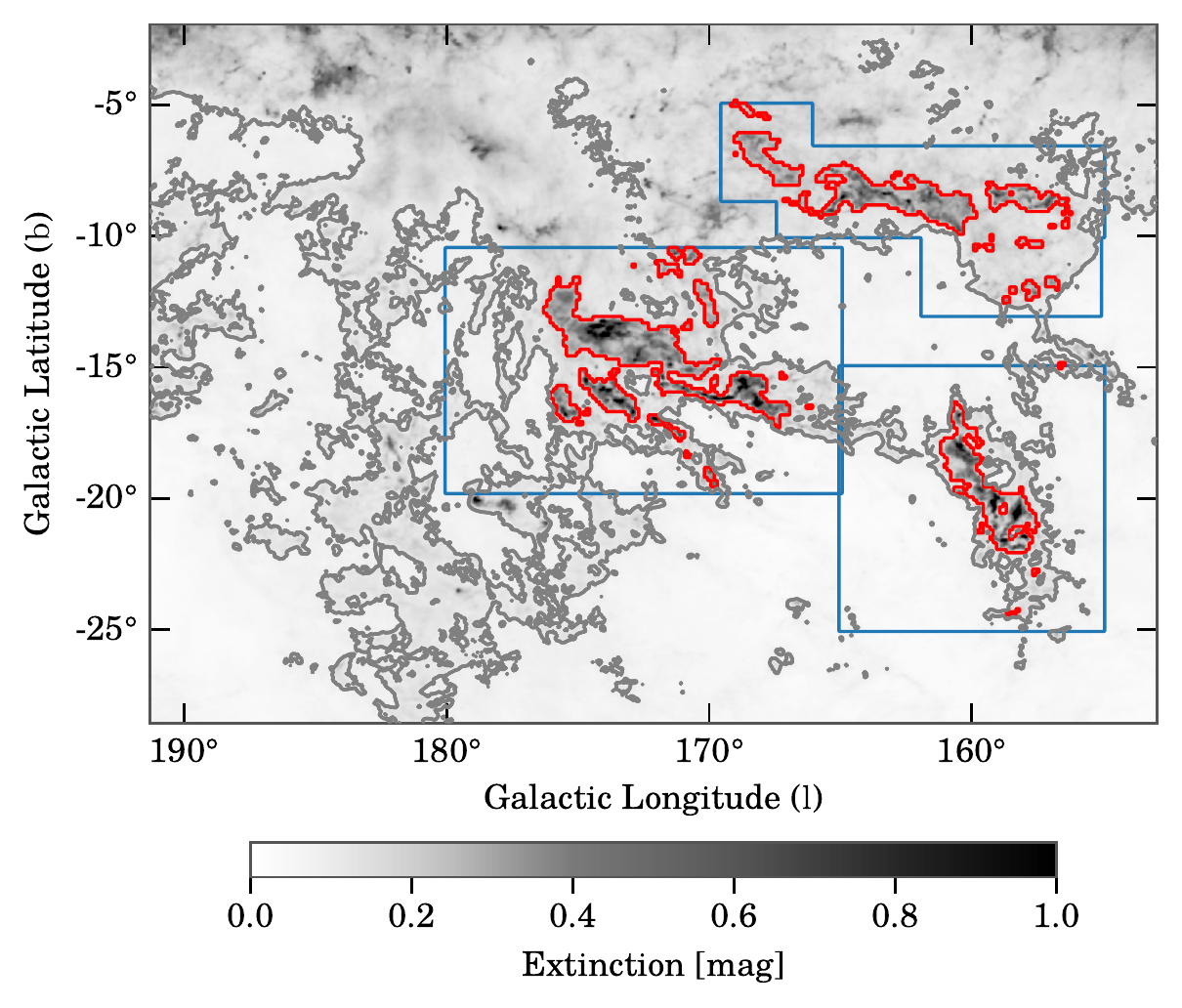}{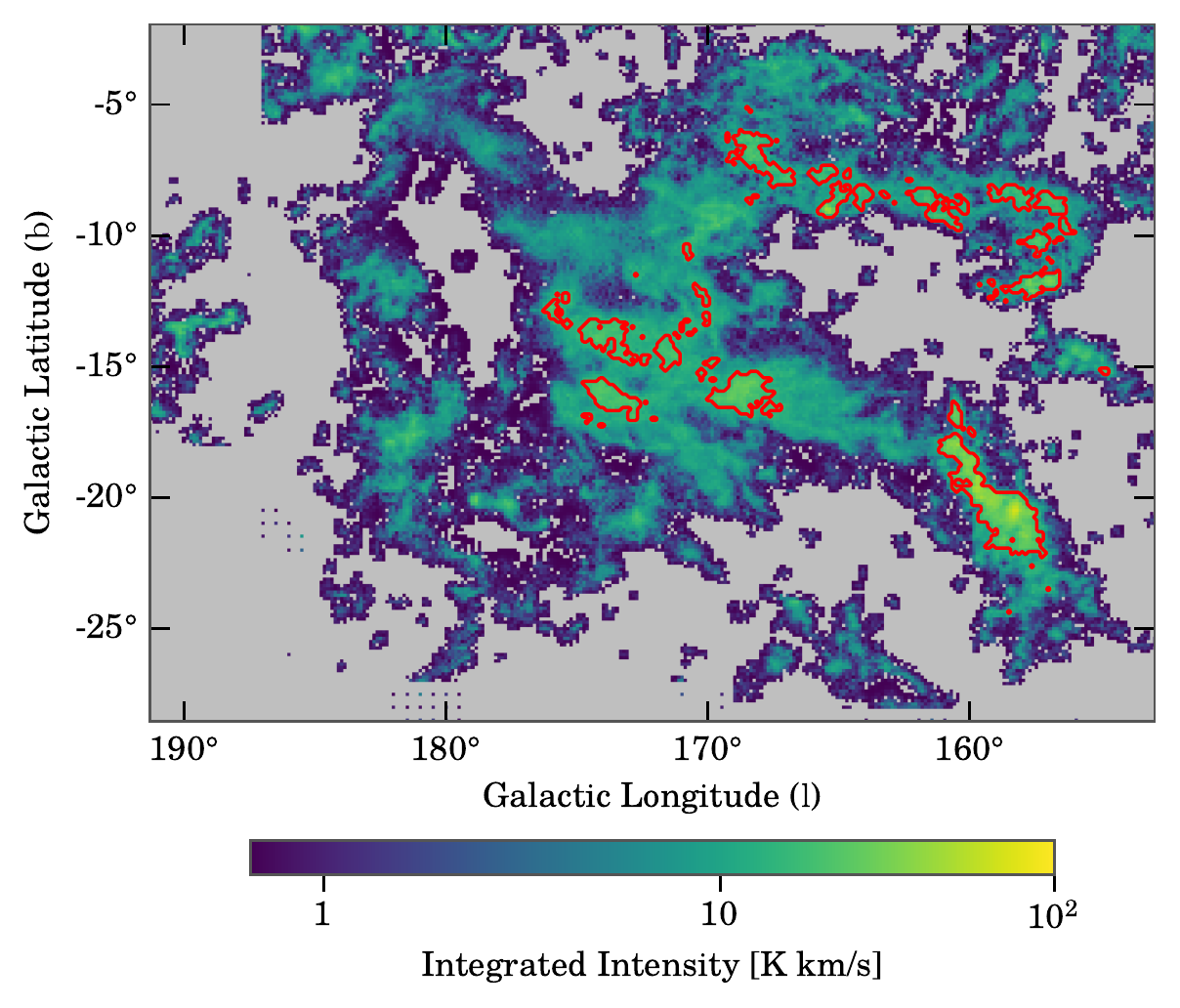}
    \caption{Taurus, Perseus, California --- }
    \label{fig:Tau_app}
    \end{figure}

\begin{figure}
    \centering
    \plottwo{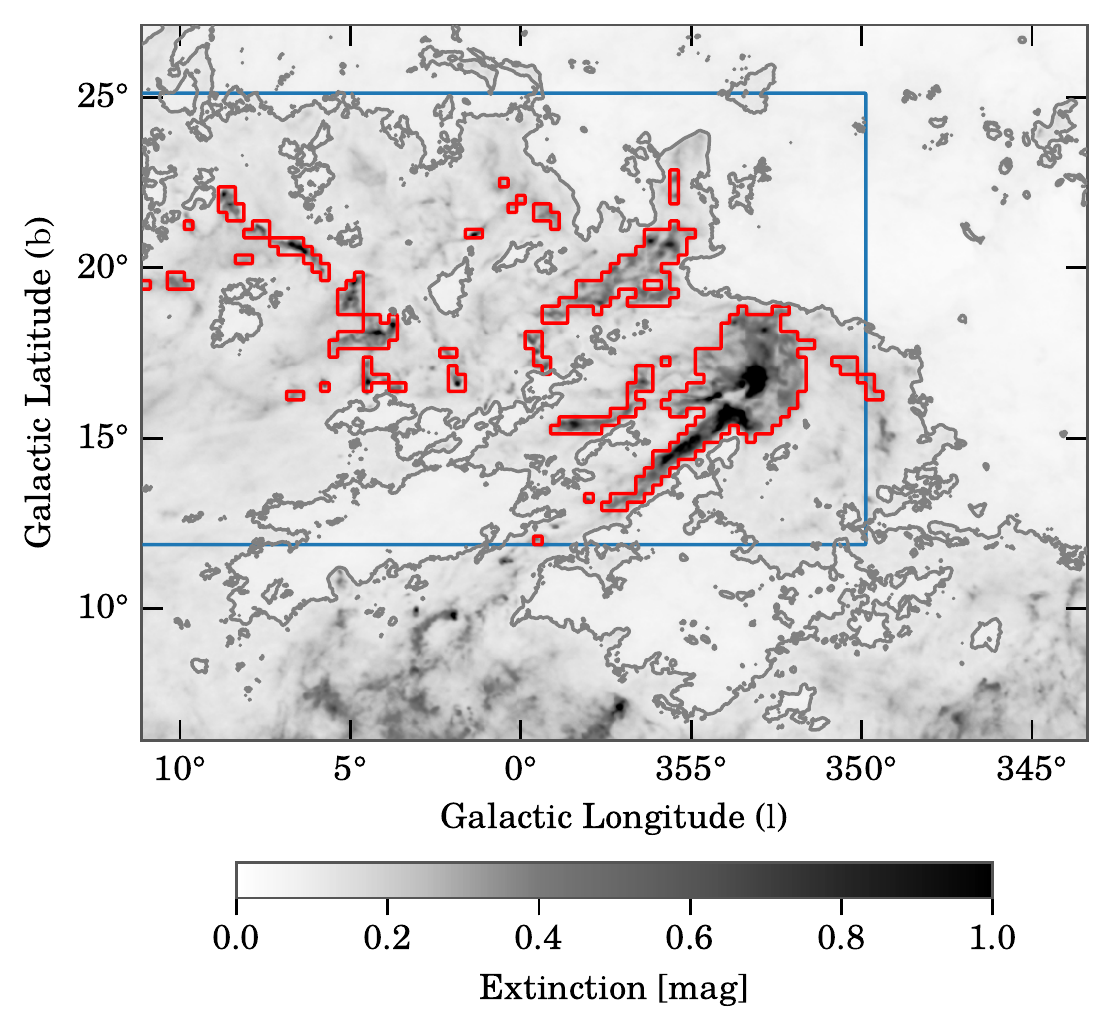}{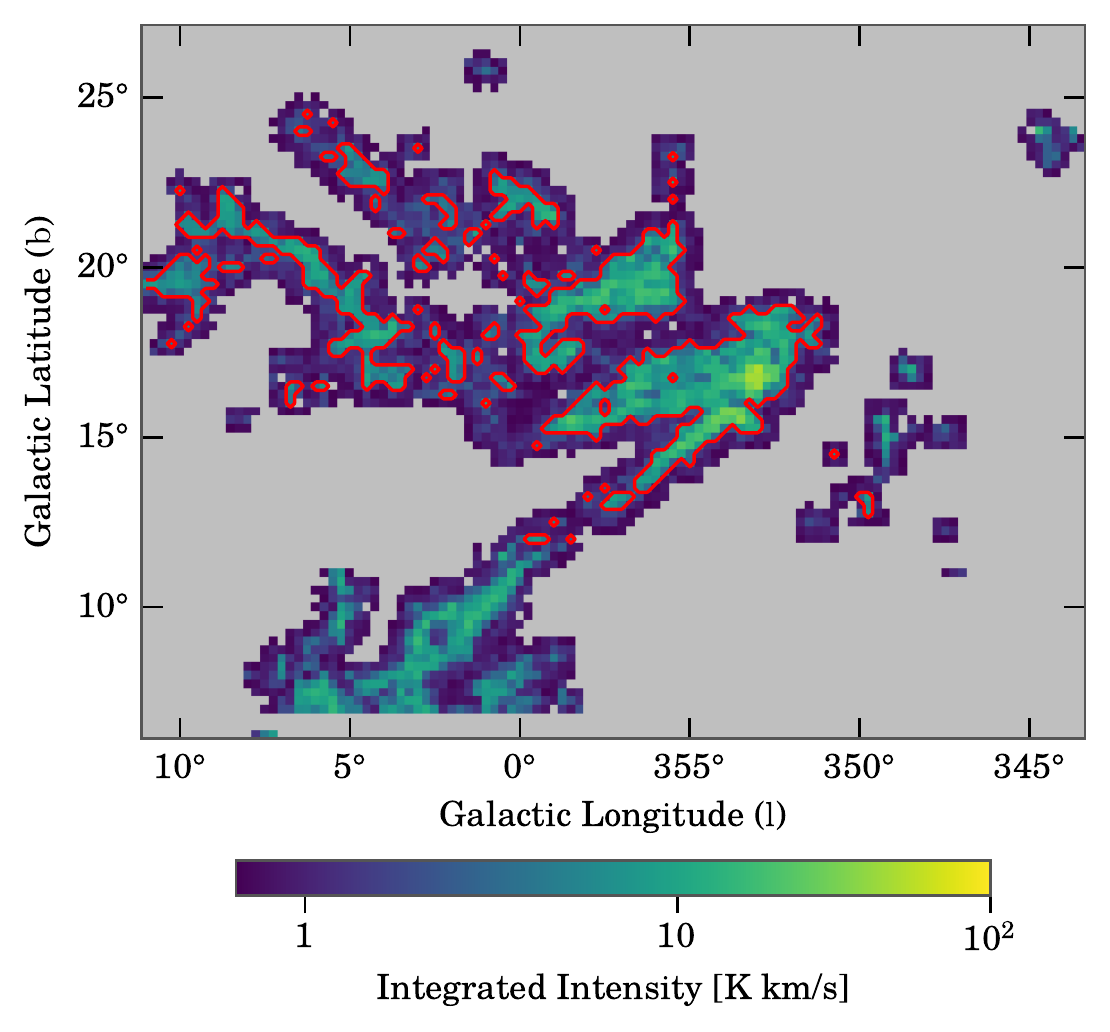}
    \caption{Ophiuchus --- }
    \label{fig:Oph_app}
    \end{figure}

\begin{figure}
    \centering
    \plottwo{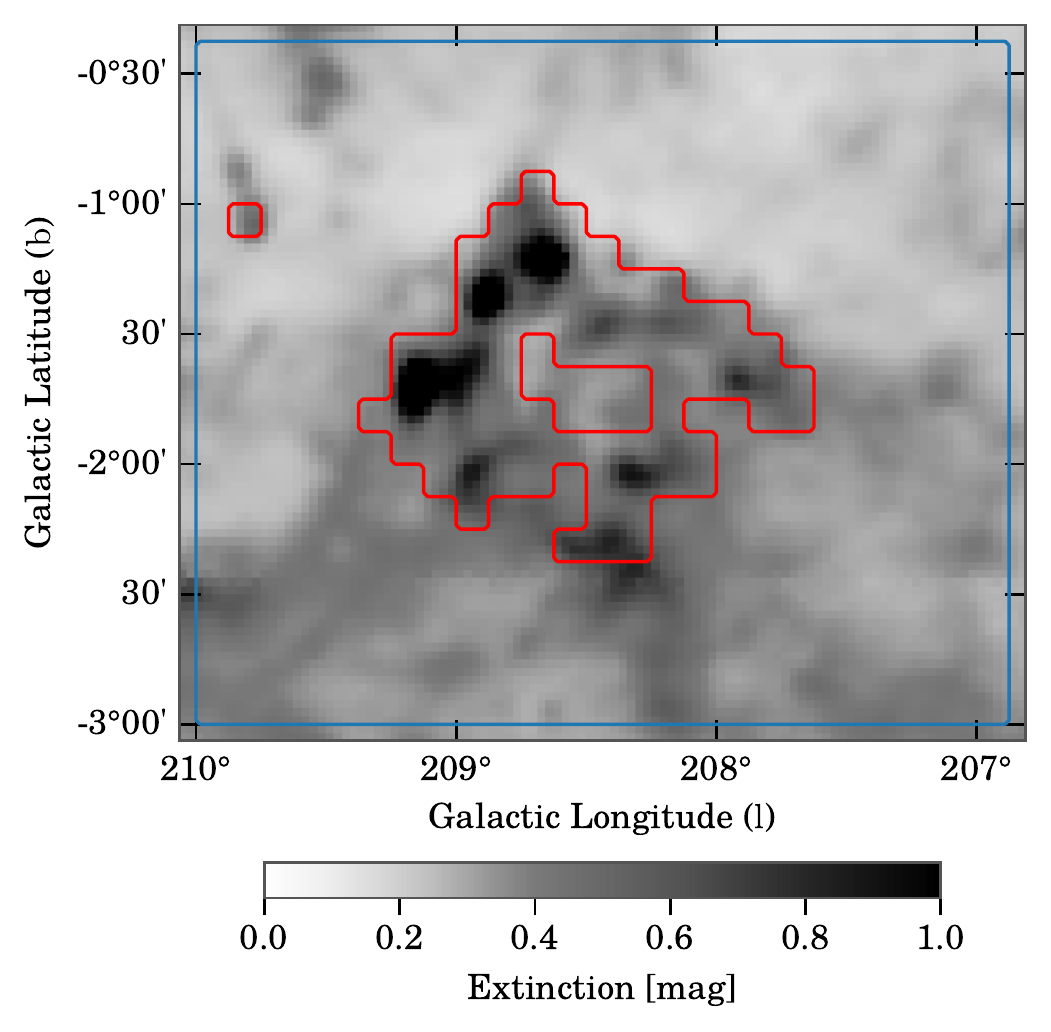}{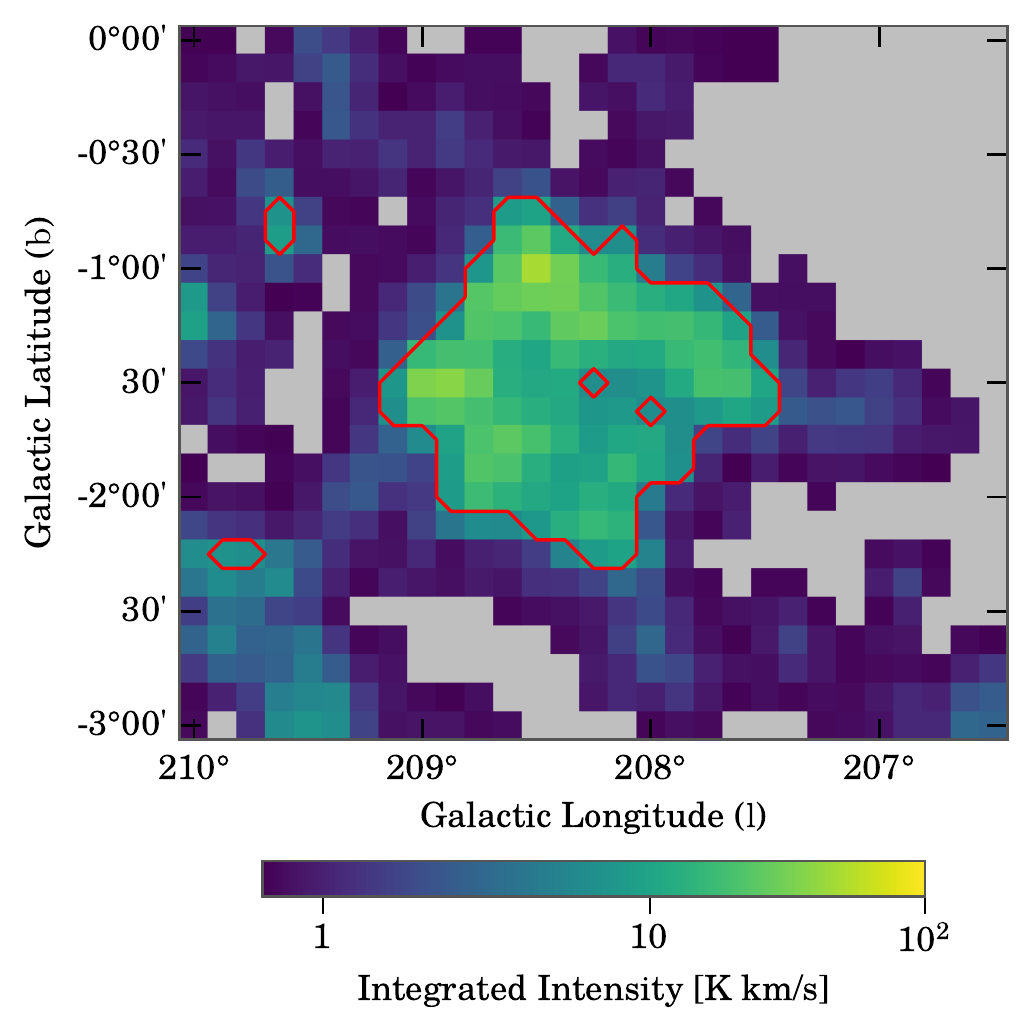}
    \caption{W3 --- }
    \label{fig:W3_app}
    \end{figure}

\begin{figure}
    \centering
    \plottwo{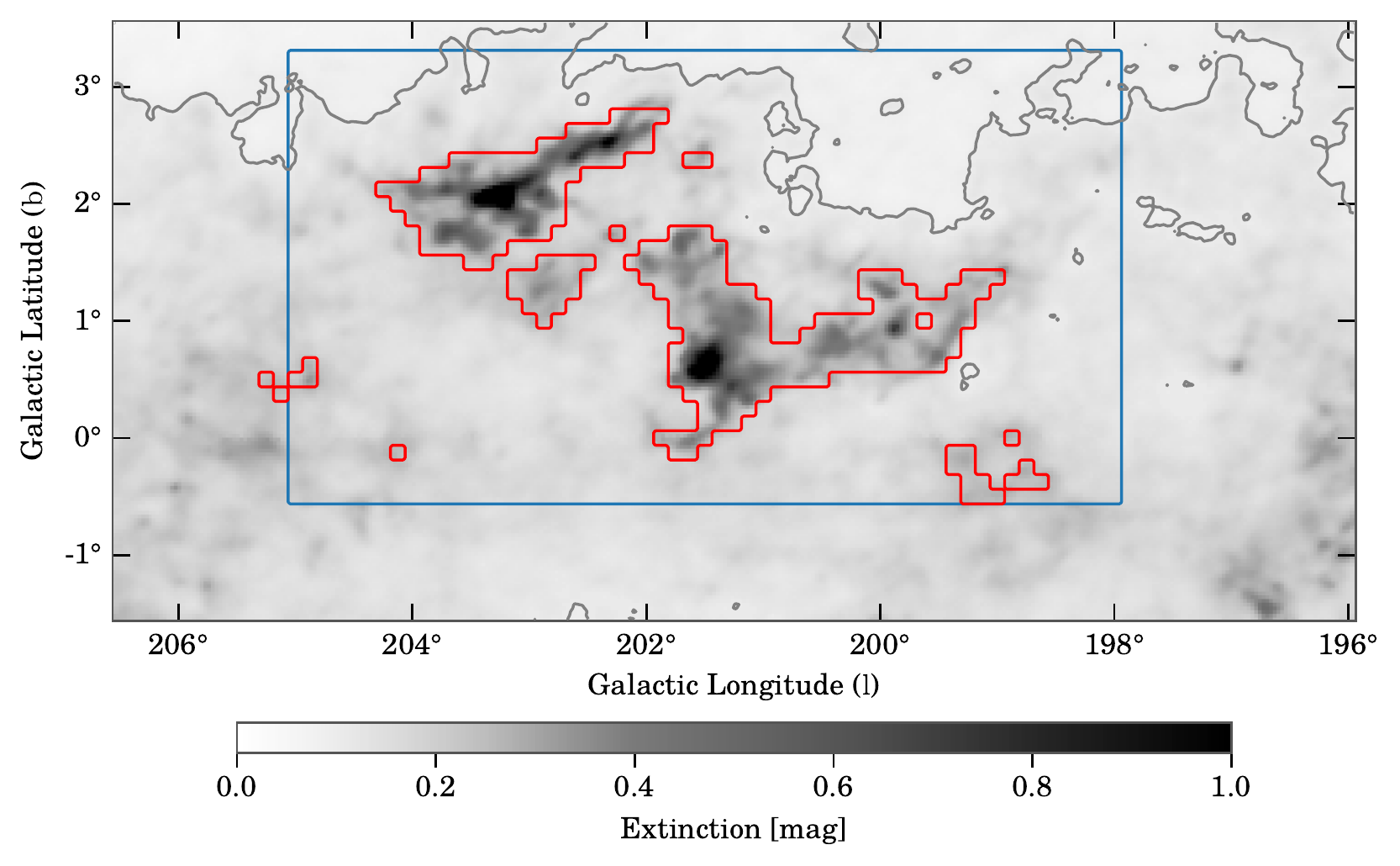}{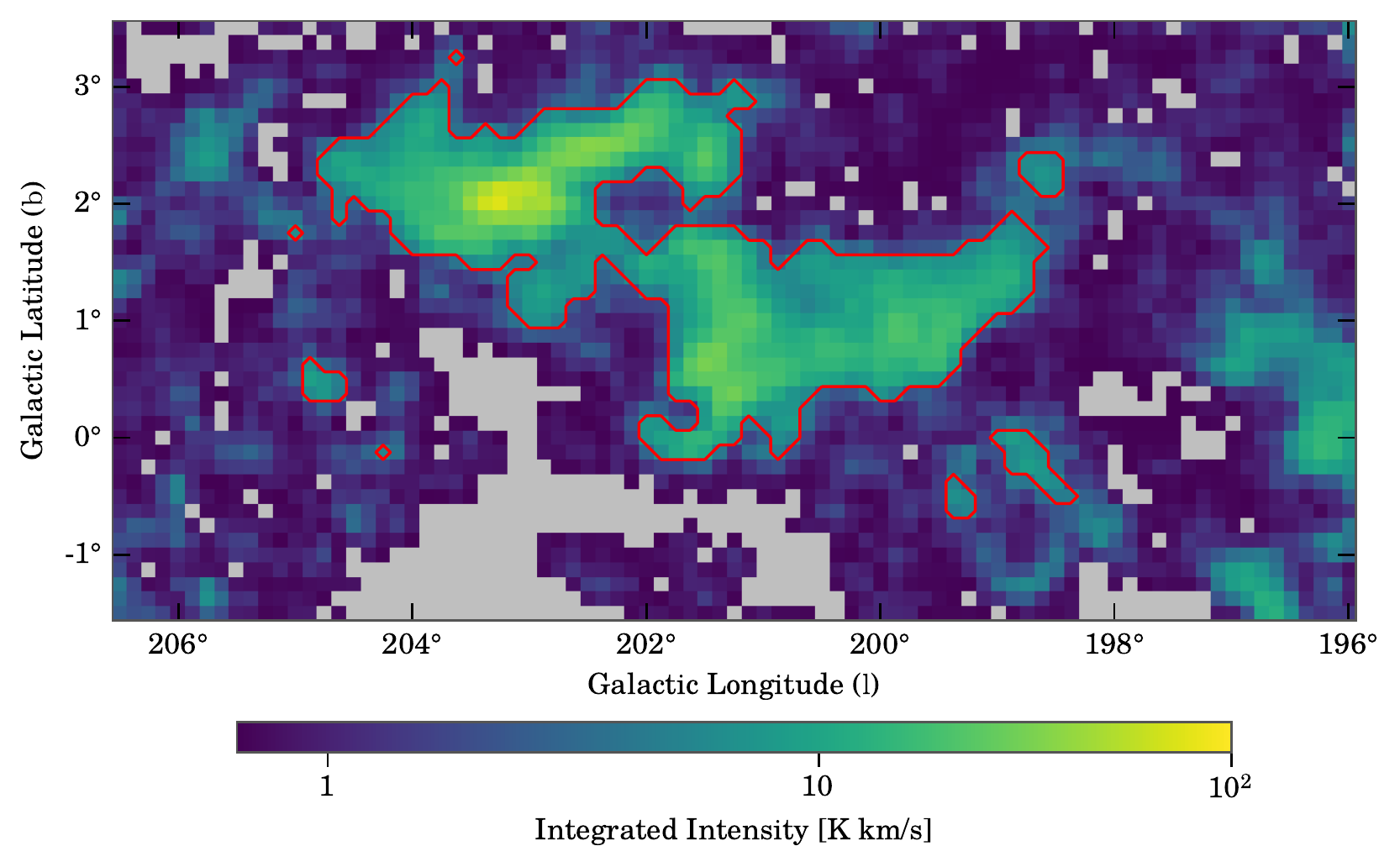}
    \caption{Mon OB1 --- }
    \label{fig:MonOB1_app}
    \end{figure}

\begin{figure}
    \centering
    \plottwo{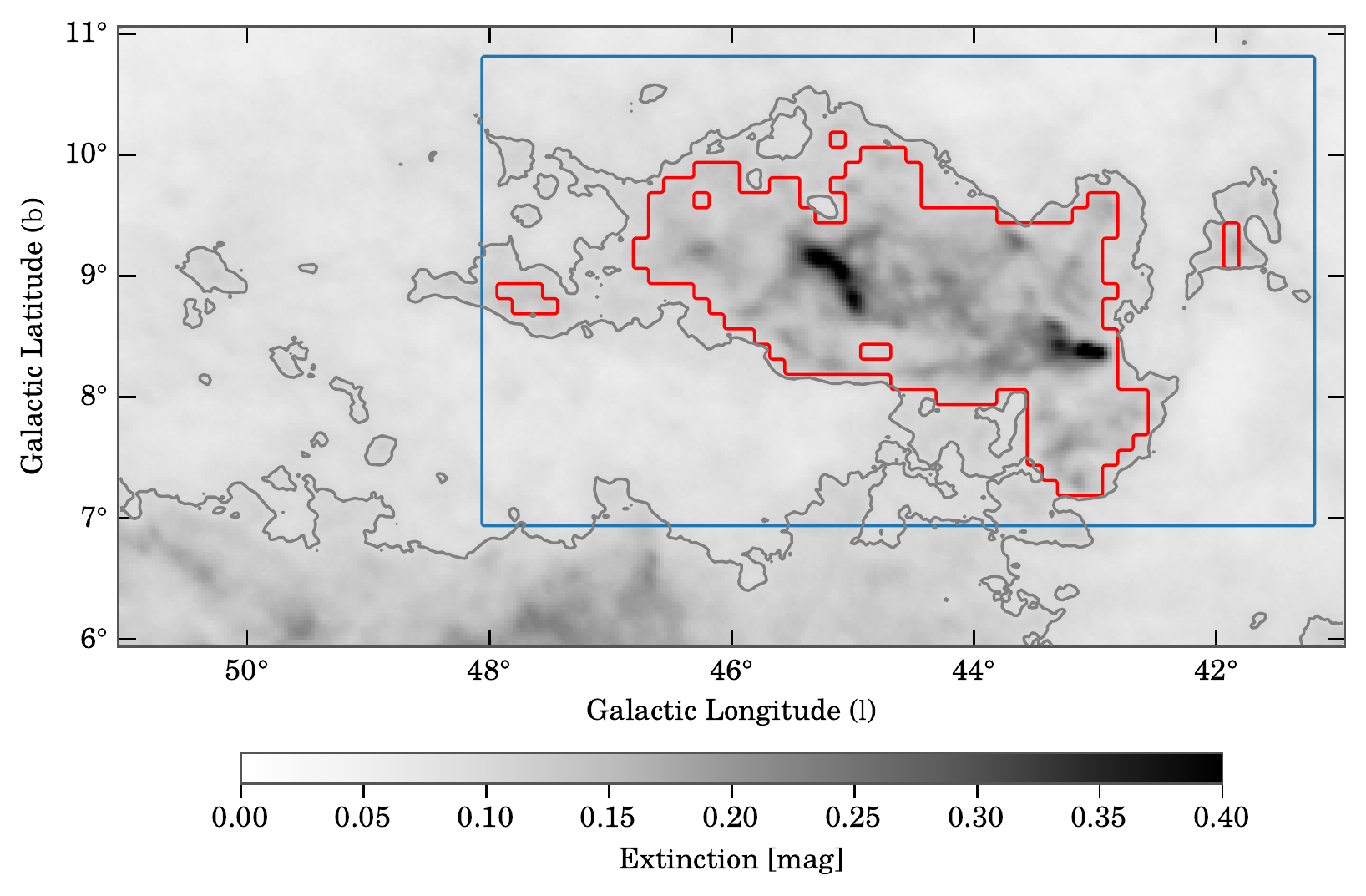}{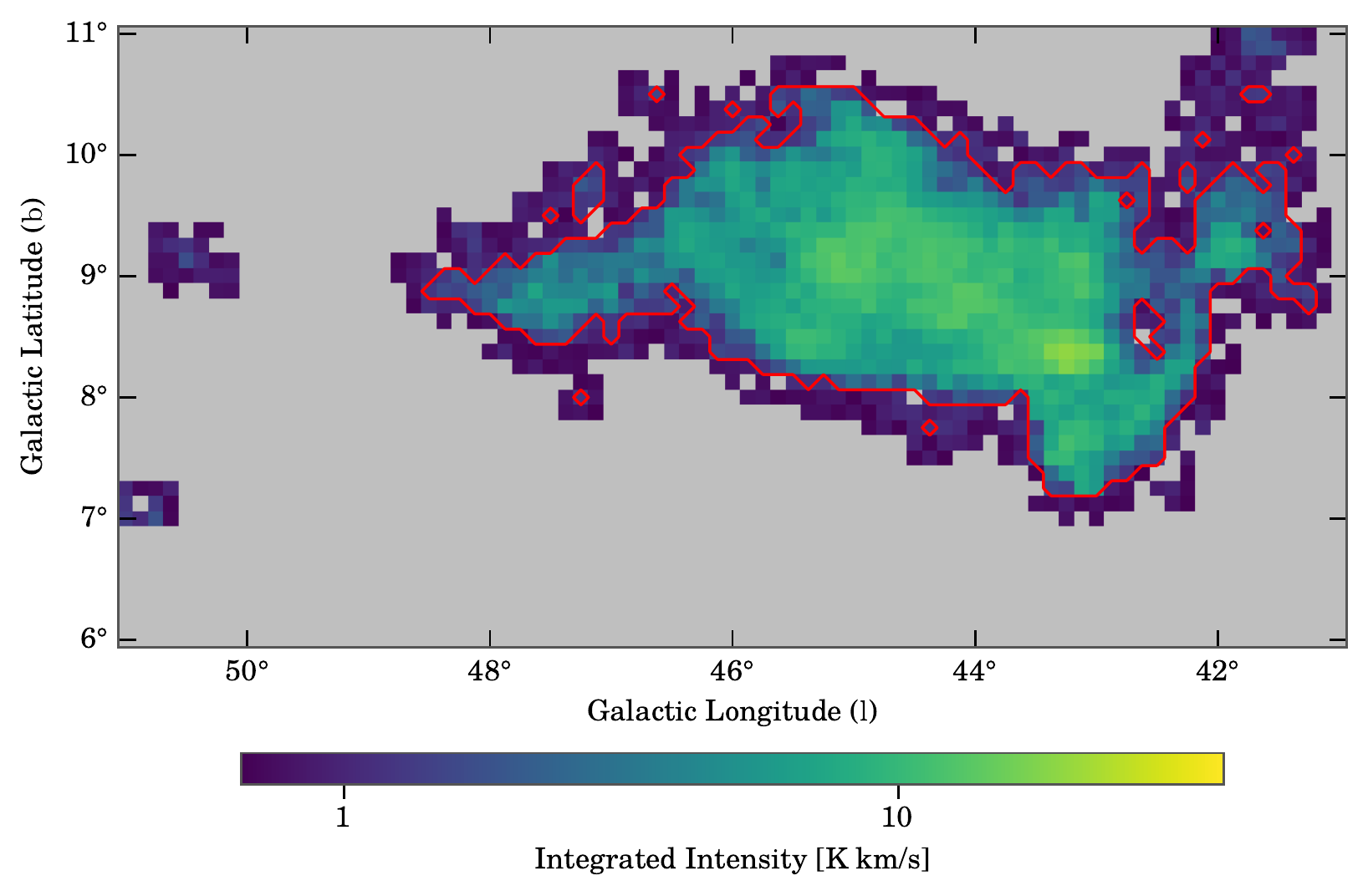}
    \caption{Hercules --- }
    \label{fig:Herc_app}
    \end{figure}

\begin{figure}
    \centering
    \plottwo{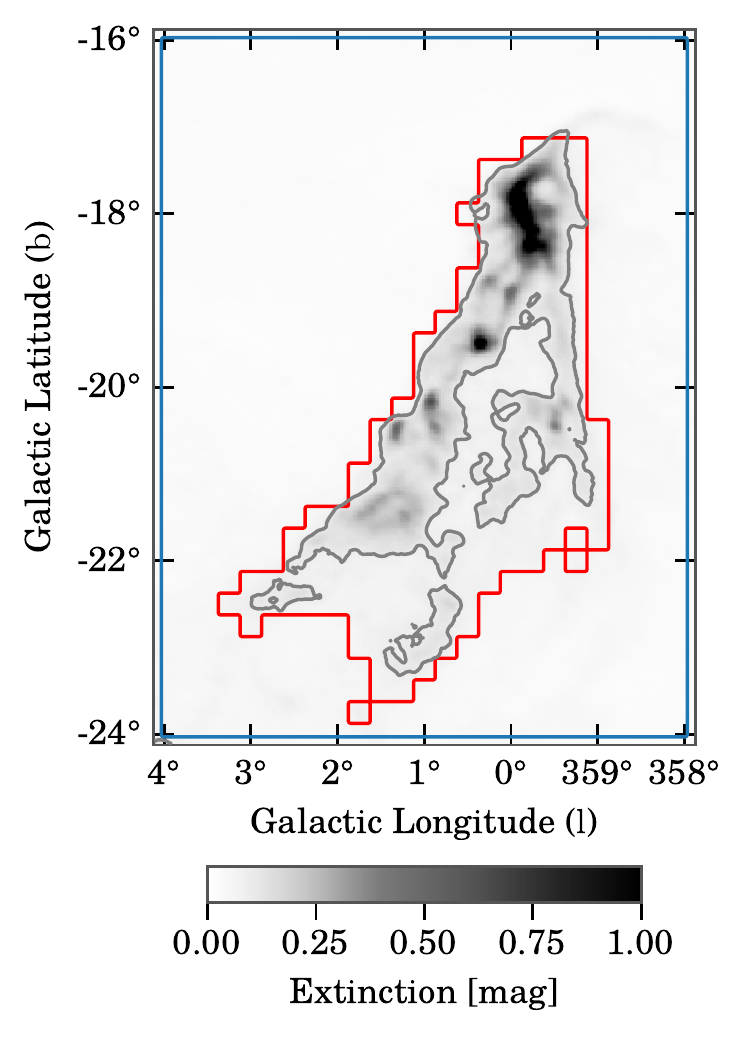}{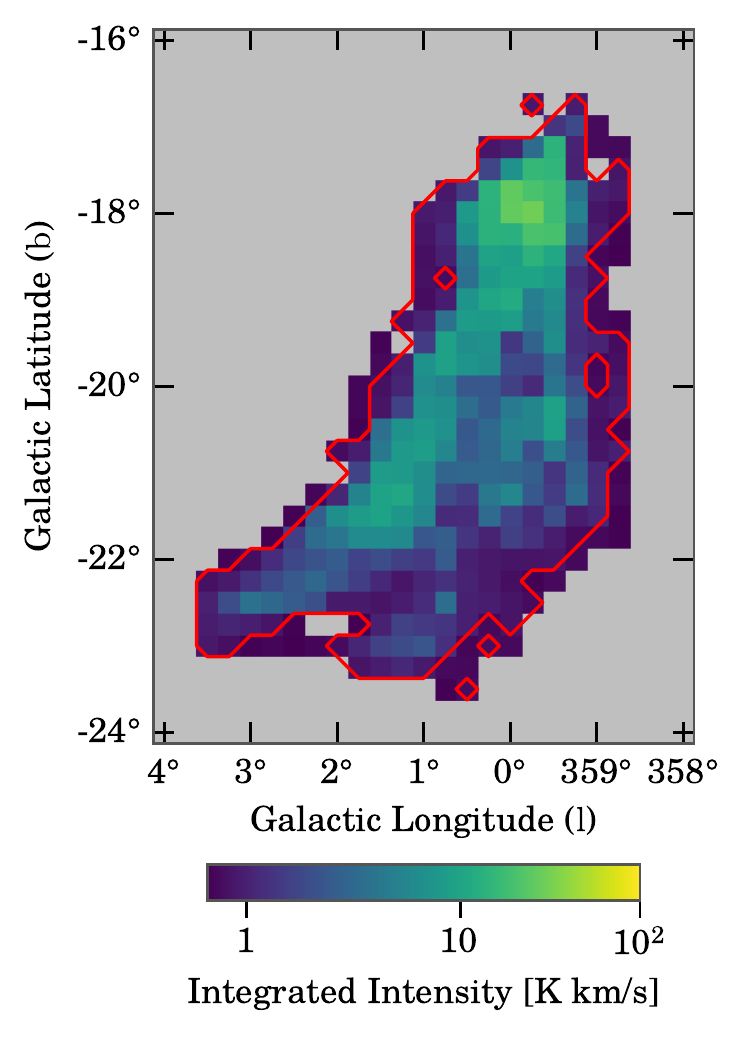}
    \caption{R Corona A --- }
    \label{fig:RCrA_app}
    \end{figure}

\begin{figure}
    \centering
    \plottwo{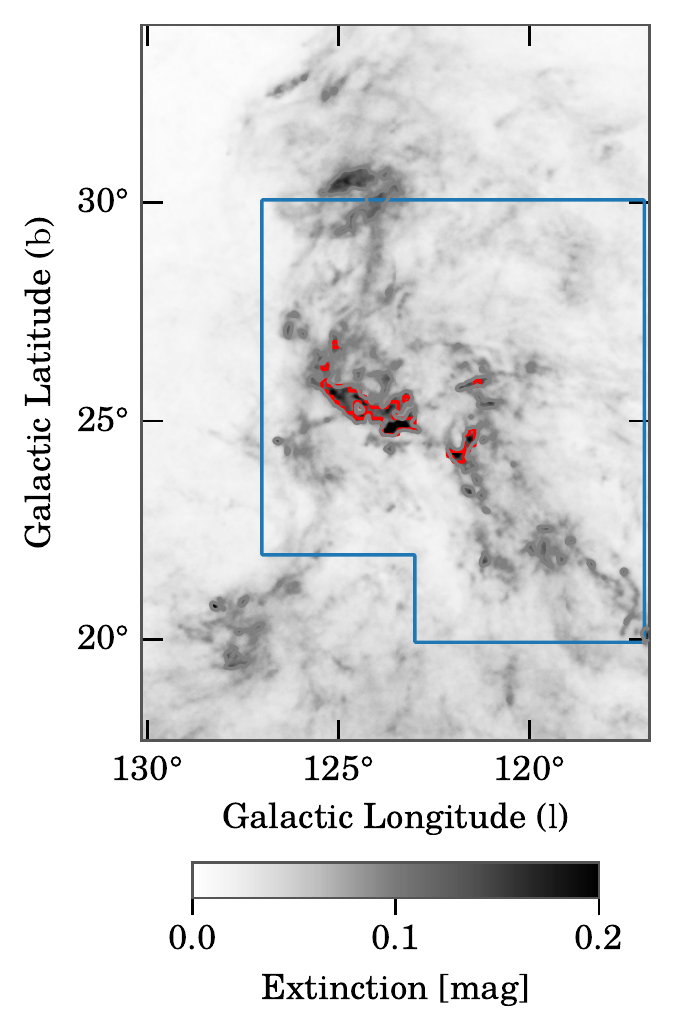}{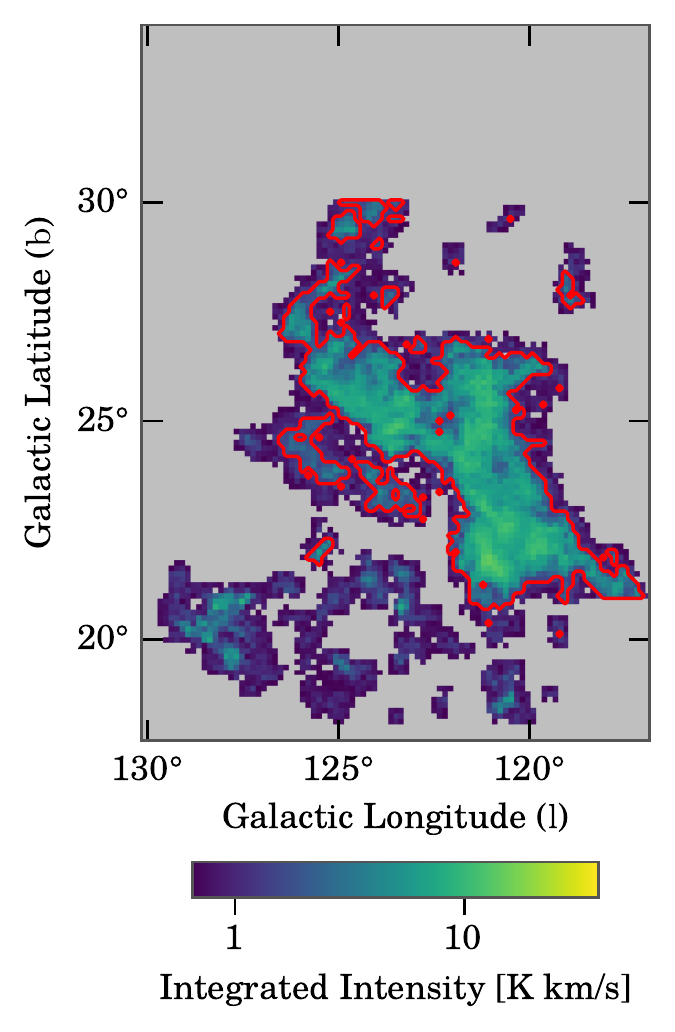}
    \caption{Polaris Flare --- }
    \label{fig:Pol_app}
    \end{figure}

In this appendix, in Figures \ref{fig:OriA_app}-\ref{fig:Pol_app}, we present maps of the \planck extinction (first) and CO emission (second) for the \citetalias{2001ApJ...547..792D} survey regions and clouds used in this study. We show the bounding box for each cloud in blue, the lowest closed extinction and \wco contours in red, and the 0.1 mag extinction contour in gray.


\end{document}